\documentclass[fleqn,usenatbib]{mnras}

\usepackage[T1]{fontenc}

\DeclareRobustCommand{\VAN}[3]{#2}
\let\VANthebibliography\thebibliography
\def\thebibliography{\DeclareRobustCommand{\VAN}[3]{##3}\VANthebibliography}

\usepackage{graphicx}	
\usepackage{amsmath}
\usepackage{physics}
\usepackage{xurl}
\usepackage{comment}
\usepackage{newtxtext,newtxmath}
\usepackage[dvipsnames]{xcolor}
\usepackage{mn2e-breakabs}
\usepackage{orcidlink}

\newcommand{\Msun}[1]{\ensuremath{\,h^{-1}\,\textup{M}_\odot}}
\newcommand{\refeq}[1]{Eq.~\eqref{#1}}
\newcommand{\reffig}[1]{Fig.~\ref{#1}}
\newcommand{\hmpc}{\,h^{-1}\,{\rm Mpc}}
\newcommand{\hgpc}{\,h^{-3}\,{\rm Gpc}^{3}}
\newcommand{\kms}{\,{\rm km}\,{\rm s}^{-1}}
\newcommand{\ergs}{\,\rm erg\,\rm s^{-1}\,{\rm cm^{-2}}}

\newcommand{\oii}{[O\,\textsc{ii}]}
\newcommand{\absum}[0]{\textsc{AbacusSummit} }

\defcitealias{mine}{Yu22}

\title[A Generalized SHAM on DESI]{The DESI One-Percent Survey: Exploring A Generalized SHAM for Multiple Tracers with the UNIT Simulation
}

\author[Jiaxi Yu et al.]
{\parbox[t]{\textwidth}{\vspace{-0.8cm}Jiaxi Yu,$^{1}$\thanks{E-mail: jiaxi.yu@epfl.ch}\orcidlink{0009-0001-7217-8006}
Cheng Zhao,$^{2,1}$\thanks{E-mail: czhao@tsinghua.edu.cn}\orcidlink{0000-0002-1991-7295}
Violeta Gonzalez-Perez,$^{3,4}$\orcidlink{0000-0001-9938-2755} 
Chia-Hsun Chuang,$^{5}$\orcidlink{0000-0002-3882-078X} 
Allyson Brodzeller,$^{5}$\orcidlink{0000-0002-8934-0954} 
Arnaud de Mattia,$^{6}$ 
Jean-Paul Kneib,$^{1,7}$ 
Alex Krolewski,$^{8,9,10}$ 
Antoine Rocher,$^{6}$\orcidlink{0000-0003-4349-6424} 
Ashley Ross,$^{11}$ 
Yunchong Wang,$^{12,13}$\orcidlink{0000-0001-8913-626X}
Sihan Yuan,$^{12,14}$\orcidlink{0000-0002-5992-7586}
Hanyu Zhang,$^{15}$\orcidlink{0000-0001-6847-5254}
Rongpu Zhou,$^{16}$\orcidlink{0000-0001-5381-4372}
Jessica Nicole Aguilar,$^{16}$ 
Steven Ahlen,$^{17}$\orcidlink{0000-0001-6098-7247} 
David Brooks,$^{18}$ 
Kyle Dawson,$^{5}$ 
Alex de la Macorra,$^{19}$\orcidlink{0000-0002-1769-1640} 
Peter Doel,$^{18}$ 
Kevin Fanning,$^{11,20}$\orcidlink{0000-0003-2371-3356} 
Andreu Font-Ribera,$^{21}$\orcidlink{0000-0002-3033-7312} 
Jaime Forero-Romero,$^{22}$\orcidlink{0000-0002-2890-3725} 
Satya Gontcho A Gontcho,$^{16}$\orcidlink{0000-0003-3142-233X} 
Klaus Honscheid,$^{11,20}$ 
Robert Kehoe,$^{23}$ 
Theodore Kisner,$^{16}$\orcidlink{0000-0003-3510-7134} 
Anthony Kremin,$^{16}$\orcidlink{0000-0001-6356-7424} 
Martin Landriau,$^{16}$\orcidlink{0000-0003-1838-8528} 
Marc Manera,$^{21,24}$\orcidlink{0000-0003-4962-8934} 
Paul Martini,$^{25,11}$\orcidlink{0000-0002-4279-4182} 
Aaron Meisner,$^{26}$\orcidlink{0000-0002-1125-7384} 
Ramon Miquel,$^{21,27}$ 
John Moustakas,$^{28}$\orcidlink{0000-0002-2733-4559} 
Jundan Nie,$^{29}$\orcidlink{0000-0001-6590-8122} 
Will Percival,$^{9,10,8}$\orcidlink{0000-0002-0644-5727} 
Claire Poppett,$^{16,30,31}$ 
Anand Raichoor,$^{16}$\orcidlink{0000-0001-5999-7923} 
Graziano Rossi,$^{32}$ 
Hee-Jong Seo,$^{33}$\orcidlink{0000-0002-6588-3508} 
Gregory Tarl\'e,$^{34}$\orcidlink{0000-0003-1704-0781} 
Zhimin Zhou,$^{29}$\orcidlink{0000-0002-4135-0977} 
Hu Zou$^{29}$\orcidlink{0000-0002-6684-3997} 
}
\vspace{0.5cm}\\
$^{1}$Laboratory of Astrophysics, \'Ecole Polytechnique F\'ed\'erale de Lausanne (EPFL), Observatoire de Sauverny, CH-1290 Versoix, Switzerland\\
$^{2}$Department of Astronomy, Tsinghua University, Beijing 100084, China\\
$^{3}$Departamento de F\'isica Te\'orica, Facultad de Ciencias, Universidad Aut\'onoma de Madrid, 28049 Madrid, Spain\\
$^{4}$Centro de Investigaci\'on Avanzada en F\'isica Fundamental (CIAFF), Facultad de Ciencias, Universidad Aut\'onoma de Madrid, 28049 Madrid, Spain\\
$^{5}$Department of Physics and Astronomy, University of Utah, Salt Lake City, UT 84112, USA\\
$^{6}$IRFU, CEA, Universit\'e Paris-Saclay, F-91191 Gif-sur-Yvette, France\\
$^{7}$Aix Marseille Universit\'e, CNRS, LAM (Laboratoire d'Astrophysique de Marseille) UMR 7326, F13388, Marseille, France\\
$^{8}$Perimeter Institute for Theoretical Physics, 31 Caroline St. North, Waterloo, ON N2L 2Y5, Canada\\
$^{9}$Waterloo Centre for Astrophysics, University of Waterloo, Waterloo, ON N2L 3G1, Canada\\
$^{10}$Department of Physics and Astronomy, University of Waterloo, 200 University Avenue W, Waterloo, ON N2L 3G1, Canada\\
$^{11}$Center for Cosmology and AstroParticle Physics, The Ohio State University, 191 West Woodruff Avenue, Columbus, OH 43210, USA\\
$^{12}$Kavli Institute for Particle Astrophysics and Cosmology, Physics Department, Stanford University, Stanford, CA 94305, USA\\
$^{14}$SLAC National Accelerator Laboratory, Menlo Park, CA 94025, USA\\
$^{15}$Department of Physics, Kansas State University, 116 Cardwell Hall, Manhattan, KS 66506, USA\\
$^{16}$Lawrence Berkeley National Laboratory, 1 Cyclotron Road, Berkeley, CA 94720, USA\\
$^{17}$Physics Dept., Boston University, 590 Commonwealth Avenue, Boston, MA 02215, USA\\
$^{18}$Department of Physics \& Astronomy, University College London, Gower Street, London, WC1E 6BT, UK\\
$^{19}$Departamento de F\'isica, Universidad de Guanajuato—DCI, C.P. 37150, Leon, Guanajuato, M\'exico\\
$^{20}$Department of Physics, The Ohio State University, 191 West Woodruff Avenue, Columbus, OH 43210, USA\\
$^{21}$Institut de F\'isica d’Altes Energies(IFAE), The Barcelona Institute of Science and Technology, Campus UAB, E-08193 Bellaterra Barcelona, Spain\\
$^{22}$Departamento de F\'isica, Universidad de los Andes, Cra. 1 No. 18A-10, Edificio Ip, CP 111711, Bogot\'a, Colombia\\
$^{23}$Department of Physics, Southern Methodist University, 3215 Daniel Avenue, Dallas, TX 75275, USA\\
$^{24}$Serra H\'{u}nter Fellow, Departament de F\'{i}sica, Universitat Au\`{o}noma de Barcelona, Bellaterra, Spain\\
$^{25}$Department of Astronomy, The Ohio State University, 140 W. 18th Ave., Columbus, OH 43210, USA\\
$^{26}$NSF’s National Optical-Infrared Astronomy Research Laboratory, 950 N. Cherry Avenue, Tucson, AZ 85719, USA\\
$^{27}$Instituci\'o Catalana de Recerca i Estudis Avan\c cats, Passeig de Llu\'is Companys, 23, E-08010 Barcelona, Spain\\
$^{28}$Department of Physics and Astronomy, Siena College, 515 Loudon Road, Loudonville, NY 12211, USA\\
$^{29}$National Astronomical Observatories, Chinese Academy of Sciences, A20 Datun Road, Chaoyang District, Beijing, 100101, People’s Republic of China\\
$^{30}$Space Sciences Laboratory, University of California, Berkeley, 7 Gauss Way, Berkeley, CA 94720, USA\\
$^{31}$University of California, Berkeley, 110 Sproul Hall\#5800 Berkeley, CA 94720, USA\\
$^{32}$Department of Physics and Astronomy, Sejong University, Seoul, 143-747, Korea\\
$^{33}$Department of Physics \& Astronomy, Ohio University, Athens, OH 45701, USA\\
$^{34}$Department of Physics, University of Michigan, Ann Arbor, MI 48109, USA\\
\vspace{-1.2cm}\\
}

\date{Accepted XXX. Received YYY; in original form ZZZ}

\pubyear{}

\begin{document}
\label{firstpage}
\pagerange{\pageref{firstpage}--\pageref{lastpage}}
\maketitle

\begin{abstract}
We perform SubHalo Abundance Matching (SHAM) studies on UNIT simulations with \{$\sigma, V_{\rm ceil}, v_{\rm smear}$\}-SHAM and \{$\sigma, V_{\rm ceil},f_{\rm sat}$\}-SHAM. They are designed to reproduce the clustering on 5--30$\,\hmpc$ of Luminous Red Galaxies (LRGs), Emission Line Galaxies (ELGs) and Quasi-Stellar Objects (QSOs) at $0.4<z<3.5$ from DESI One Percent Survey. $V_{\rm ceil}$ is the incompleteness of the massive host (sub)haloes and is the key to the generalized SHAM. $v_{\rm smear}$ models the clustering effect of redshift uncertainties, providing measurments consistent with those from repeat observations. A free satellite fraction $f_{\rm sat}$ is necessary to reproduce the clustering of ELGs. We find ELGs present a more complex galaxy--halo mass relation than LRGs reflected in their weak constraints on $\sigma$. LRGs, QSOs and ELGs show increasing $V_{\rm ceil}$ values, corresponding to the massive galaxy incompleteness of LRGs, the quenched star formation of ELGs and the quenched black hole accretion of QSOs. For LRGs, a Gaussian $v_{\rm smear}$ presents a better profile for sub-samples at redshift bins than a Lorentzian profile used for other tracers. The impact of the statistical redshift uncertainty on ELG clustering is negligible. The best-fitting satellite fraction for DESI ELGs is around 4 per cent, lower than previous estimations for ELGs. The mean halo mass log$_{10}(\langle M_{\rm vir}\rangle)$ in $\Msun{}$ for LRGs, ELGs and QSOs are ${13.16\pm0.01}$, ${11.90\pm0.06}$ and ${12.66\pm0.45}$ respectively. Our generalized SHAM algorithms facilitate the production of mult-tracer galaxy mocks for cosmological tests. 

\end{abstract}



\section{Introduction}
$\Lambda$CDM is the standard model of modern cosmology that describes the evolution of the Universe. In this framework, two dark components, dark matter and dark energy, comprise 95 per cent of the total energy density. The nature of dark energy can be explored using baryonic acoustic oscillation \citep[BAO;][]{bao1998}, a standard ruler of the universe. Meanwhile, redshift-space distortion \citep[RSD,][]{Kaiser1987} embodies the growth rate of the large-scale structure (LSS) which is dominated by the evolution of dark matter. 

In observation, BAO and RSD can be measured by spectroscopic galaxy surveys that observe millions of spectra of galaxies and QSOs. 
With BAO and RSD, the precision of cosmological parameters, such as $\Omega_{\rm \Lambda}$, $H_0$ and $\sigma_8$, have reached per cent level \citep{Alam20}, constrained with data from the Baryon Oscillation Spectroscopic Survey \citep[BOSS, 2008--2014;][]{Dawson2013}, and the extended BOSS \citep[eBOSS, 2014--2020;][]{Dawson2015} in the Sloan Sky Digital Survey\footnote{\url{http://www.sdss.org/}} \citep[SDSS;][]{SDSSIII,SDSSIV}.
BOSS and eBOSS have probed 1,547,553 luminous red galaxies (LRGs) at $0.2<z<1.0$, 173,736 emission line galaxies (ELGs) at $0.6<z<1.1$, and 343,708 QSOs at $0.8<z<2.2$ \citep{2021PhRvD.103h3533A}.

The largest ongoing spectroscopic survey, the Dark Energy Spectroscopic Instrument \citep[DESI, 2021--2026;][]{DESICollaboration2016a,DESICollaboration2016b} is a robotic, fiber-fed, highly multiplexed survey that operates on the Mayall 4-meter telescope at Kitt Peak National Observatory. It aims to explore the nature of dark energy via the most precise measurement of the 3D Universe in $14,000\,\rm deg^2$ of the sky after 5 years of observations \citep{Levi2013}. The list of targets to be observed by DESI \citep{2023DESItarget} is determined with the help of the imaging from the DESI Legacy Imaging Surveys (\citet{BASS2017,DES2019}; \textcolor{blue}{Schlegel et al. in prep.}).
The preliminary selection of targets was done in 2020 for the Milky Way Survey \citep[MWS;][]{preTS_MWS2020}, Bright Galaxy Survey \citep[BGS;][]{preTS_BGS2020}, LRG \citep{preTS_LRG2020}, ELG \citep{preTS_ELG2020}, QSO \citep{preTS_QSO2020}.

DESI started its first light observation in 2020 and will make public its Early Data Release \citep[EDR;][]{sv} and the Siena Galaxy Atlas (SGA; \textcolor{blue}{Moustakas et al. in prep.})
in 2023. EDR contains LSS catalogues that include redshift measurements, their corresponding random catalogues, and the clustering output (\citet{edr}; \textcolor{blue}{Lasker et al. in prep.}).
The One Percent Survey is a part of the EDR. It has covered around one per cent of the 5-year sky footprint and observed more than 90,000 LRGs, 270,000 ELGs, 30,000 QSOs, and 150,000 low-redshift galaxies (Bright Galaxy Sample, BGS) \citep{sv}. Despite the smaller numbers of galaxies and QSOs compared to the SDSS data, the number densities of tracers from the DESI One Percent Survey are larger than those of BOSS and eBOSS (introduced later in Table~\ref{tab:3param result}). Additionally, the rate between the observed targets and the targets of the One-Percent Surveys is larger than 85 per cent for all tracers (\citet{edr}; \textcolor{blue}{Lasker et al. in prep.}). Thus, data from the One Percent Survey are sufficient for small-scale clustering analysis, such as the galaxy--halo connection study.

The relationship between haloes and galaxies is crucial for the modelling of galaxy clustering. However, this relation is highly non-linear and generally subject to local environmental effects \citep[e.g.,][]{local_gh2011,local_gh2012}. 
SubHalo Abundance Matching \citep[SHAM,][]{Kravtsov2004,tasitsiomi_modeling_2004,Conroy2006,Behroozi2010} is an intuitive empirical method to model this non-linear relation based on $N$-body simulations that resolve hierarchical structures, including both haloes and subhaloes. 
This method assigns the most massive or brightest galaxy to centres of the most massive haloes in the case of central galaxies and subhaloes for satellite galaxies. 
The resulting probability of hosting a central/satellite galaxy in a halo/subhalo is a function of their (sub)halo mass, $P(M_{\rm halo})$. The shape of this probability is related to the stellar properties determined empirically.

As clustering observations and simulations improve, more advanced versions of SHAM algorithms are developed. For instance, \cite{tasitsiomi_modeling_2004} introduced a Gaussian scattering with dispersion $\sigma$ to the halo mass, to model the Gaussian residual in the galaxy--halo mass relation \citep[e.g.,][]{willick_homogeneous_1997,steinmetz_cosmological_1999}. \citet{trujillo-gomez_galaxies_2011} and \citet{Reddick2013} proposed using the peak maximum circular velocity, $V_{\rm peak}$, instead of the halo mass, $M_{\rm vir}$, as it is closely associated with stellar mass and it is immune to the tidal stripping of subhaloes and pseudo evolution of their $M_{\rm vir}$. \cite{favole_clustering_2016,QSOfsat_2017} proposed a SHAM implementation that takes into account the fact that ELGs and QSOs are incomplete in the massive stellar-mass end. There are also SHAM variants that make use of secondary halo/galaxy properties \citep[e.g.,][]{Hearin_SHAM_2013, Ginevra_SHAM_2022}. SHAM methods can also include assembly bias and orphan galaxies \citep{assemblybias_2017, Behroozi2019, SHAMe2021, Joe2022}. 

In this work, we use two SHAM implementations that are essentially variants of one algorithm: $\{\sigma,V_{\rm ceil},v_{\rm smear}\}$-SHAM ($v_{\rm smear}$-SHAM hereafter) and $\{\sigma,V_{\rm ceil},f_{\rm sat}\}$-SHAM ($f_{\rm sat}$-SHAM hereafter). The $v_{\rm smear}$-SHAM was used to study BOSS/eBOSS LRGs \citep[][\citetalias{mine} hereafter]{mine} and here we use it to model LRGs and QSOs. Here, we introduce the $f_{\rm sat}$-SHAM to be able to correctly reproduce the clustering of DESI ELGs. 
The free parameters in these SHAMs model the following aspects: the scatter in the galaxy--halo mass relation, $\sigma$; an upper limit of $\sigma$-scattered $V_{\rm peak}$ set by $V_{\rm ceil}$ (in percentage), which reduces the possibility of massive (sub)haloes hosting a given type of galaxy or QSO; the uncertainty in spectroscopic redshift determination, $v_{\rm smear}$; and the fraction of satellite galaxies, $f_{\rm sat}$, which we find to be only needed for reproducing the clustering of ELGs. 

This paper is arranged as follows. In Section~\ref{sec:data} we describe the early data release of DESI, including repeat observations and statistical redshift uncertainty measurements, the UNIT $N$-body simulation, and the covariance matrix. The SHAM implementation and fitting are introduced in Section~\ref{sec:method}. In Section~\ref{sec:results}, we present the best-fitting results of SHAM and the interpretations of parameters for different tracers. We conclude our findings in Section~\ref{sec:conclusion}. 

This paper is one of the first series papers from the DESI galaxy--halo connection topical group. Papers released with EDR for One Percent Survey analysis that utilise \absum \citep{AbacusSummit} simulations are \citet{Yuan2023} for LRG and QSO HOD, \citet{Rocher2023} for ELG HOD. \citet{Prada2023} is an overview for SHAM based on \textsc{Uchuu} \citep{2021Ishiyama}. 
A stellar-mass-split abundance matching applied on \textsc{CosmicGrowth} \citep{2019SCPMA..6219511J} is also used to study DESI LRG--ELG cross-correlations \citep{Gao2023}. Other works will be published along with later data releases. 

\begin{figure*}
    \includegraphics[width=\linewidth,scale=0.8]{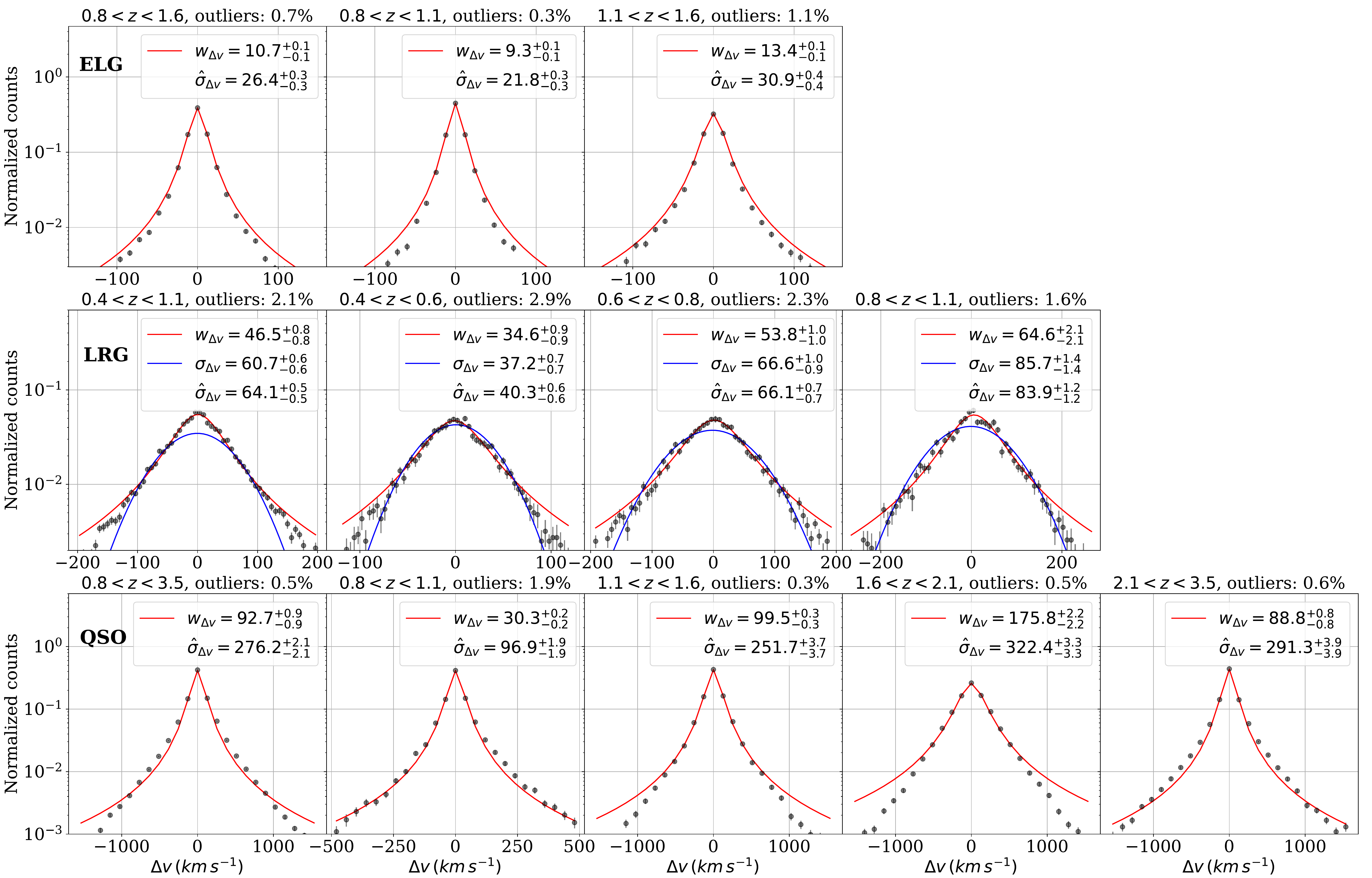}
    \caption{The statistical redshift uncertainty estimated with the histogram of the redshift difference (black filled circles with error bars) from repeat observation taken during the early stage of Survey Validation. The first, second and third rows are ELGs, LRGs and QSOs respectively. The first columns of all rows are results for total samples, while the rest are for sub-samples at redshift bins. The statistical redshift uncertainty measured by Lorentzian functions $w_{\Delta v}$ (solid red lines) and standard deviations $\hat{\sigma}_{\Delta v}$ of $\Delta v$ is presented in the label of each subplot. For LRG samples, we also fit $\Delta v$ histograms with Gaussian functions (blue solid lines), providing their best-fitting dispersion $\sigma_{\Delta v}$ in the label as well. The fraction of $\Delta v$ that are not included in the fittings is indicated as outlier fractions in titles.}
    \label{fig:deltav measurement}
\end{figure*}

\section{Data}
\label{sec:data}
\subsection{DESI Early Data Release}
\label{subsec:EDR}
DESI, a 5-year spectroscopic survey, started instrumental tests in 2020 to ensure the 5000 fibres controlled by the robotic positioners could work properly in the focal plane over a 3-degree field of view \citep{2022DESIinstrument, DESIfocalplane,DESIcorrector2023}. After the commissioning, DESI conducted its Survey Validation campaign \citep{sv}. It aims to validate the spectra reduction pipeline \citep{DESIspec}, assess the quality of data provided by \textsc{Redrock}\footnote{\url{https://github.com/desihub/redrock}} that derives the target type and redshift from spectra (\textcolor{blue}{Bailey et al. in prep.}), 
and optimise the target selection \citep{DESIsops} and fibre assignment (\textcolor{blue}{Raichoor et al. in prep.})
of DESI. During the campaign, it explores target selection criteria broader than those of the 5-year main survey and observes objects typically four times longer than the main survey; in addition, to perform the visual inspection, few tiles are observed approximately 10 times longer than in the main survey \citep{QSO_VI,DESIVI}. For this reason, there are many repeat observations for each object. Later in April and May of 2021, DESI observed its One Percent Survey that covered about 1 per cent of the footprint of the 5-year main survey and used target selection criteria close to those of the main survey \citep{MWS,BGS,DESI_LRG_TS, DESI_ELG_TS, DESI_QSO_TS}. The observation field is composed of 20 non-overlapping rosettes, each observed at least 12 times. This ensures very high fibre assignment completeness (larger than 85 per cent for ELGs and over 94 per cent for the rest of the tracers) in this region (\citet{sv,edr}; \textcolor{blue}{Lasker et al. in prep.}). As there are more exposures for objects that do not have a reliable redshift after the first observation, data from the One Percent Survey have a high redshift-success rate. 

Our SHAM method aims to reproduce the clustering of LRGs at $0.4<z<1.1$, ELGs at $0.8<z<1.6$, and QSOs at $0.8<z<3.5$ from the One Percent Survey in the range 5--30$\hmpc$. LRG samples are divided into three smaller redshift ranges: $0.4<z<0.6$, $0.6<z<0.8$, $0.8<z<1.1$. ELG samples are divided into two redshift bins: $0.8<z<1.1$ and $1.1<z<1.6$. QSOs are observed at $0.8<z<3.5$, and are divided into $0.8<z<1.1$, $1.1<z<1.6$, $1.6<z<2.1$, $2.1<z<3.5$. 



\subsubsection{Galaxy Weights}
\label{subsubsec:weights}
To obtain an unbiased measurement in the galaxy clustering, we employ the FKP weight $w_{\rm FKP}$, the pairwise-inverse-probability (PIP) weight \citep{bianchi17}, 
and the angular-up weight (ANG) \citep{percival17} for pairs of galaxies. In the calculation of the effective redshift, we use the total weight $w_{\rm tot}$ 
\begin{equation}
    w_{\rm tot} = w_{\rm FKP}w_{\rm comp}.
    \label{eq:wtot}
\end{equation}
where $w_{\rm comp}$ is the fibre-assignment completeness weight provided in the LSS catalogue. We briefly describe all of them below and we refer the readers to \citet{edr}; \textcolor{blue}{Lasker et al. in prep.} for more details.

The One Percent Survey LSS catalogues provide $w_{\rm FKP}$ \citep{FKP1994} that minimises the variance in the clustering estimator (see Section~\ref{subsec:2pcf}) when the observed number density of tracers varies with redshift 
\begin{equation}
    w_{\rm FKP} = \frac{1}{1+\overline{n}(z)P_0},
\end{equation}
where $\overline{n}(z)$ is the average number density at redshift $z$, and $P_0$ is the amplitude of the observed power spectrum at $k\approx {0.15}\, h\,{\rm Mpc}^{-1}$. $P_0=10000, 4000, 6000\, h^{-3}\,{\rm Mpc}^{3}$ for LRGs, ELGs and QSOs respectively \citep{edr}. 

The PIP+ANG weighting scheme has been developed to correct the missing galaxy pairs due to fibre collision. 
\citet{Mohammad_2020} have proved that PIP+ANG weights provide an unbiased clustering down to 0.1$\hmpc$. So the clustering measurement provided by the EDR has implemented this weighting scheme in addition to $w_{\rm FKP}$ (Section~\ref{subsec:2pcf}). The $w_{\rm comp}$ provided in the LSS catalogues is for correcting the observational incompleteness due to the fibre assignment \citep{edr}. PIP and ANG weights, as well as $w_{\rm comp}$, are all calculated with simulations of fibre assignment as described in \textcolor{blue}{Lasker et al. in prep.}. 

We calculate the effective redshift of pairs of galaxies at redshifts $z_i$ and $z_j$ \citep[e.g., ][]{eBOSS_LRG}, with
\begin{equation}
z_{\rm eff} = \frac{\sum_{i,j}w_{\text{tot},i}w_{\text{tot},j}(z_i+z_j)/2}{\sum_{i,j}w_{\text{tot},i}w_{\text{tot},j}},
\label{eq:zeff}
\end{equation}
The effective volume $V_{\rm eff, obs}$ \citep{2013Veff} also involves $P_0$ as 
\begin{equation}
    V_{\rm eff} = \sum_i \left(\frac{\bar{n}(z_i)P_0}{1+\bar{n}(z_i)P_0}\right)^2 \Delta V(z_i),
    \label{eq:Veff}
\end{equation}
where $\Delta V(z_i)$ is the comoving survey volume and $\overline{n}(z_i)$ is the mean number density of the tracer inside the redshift bin $z_i$. The effective number density is calculated as
\begin{equation}
\label{eq:neff}
\begin{split}
    n_{\rm eff} &=  \sqrt{ \frac{\sum_i \bar{n}(z_i)^2 \Delta V(z_i)} { \sum_i\Delta V(z_i)}}
\end{split}
\end{equation}
We present in Table~\ref{tab:3param result} all the tracers we use, their redshift ranges and the corresponding effective redshifts $z_{\rm eff}$, the effective volume $V_{\rm eff}$ and the number density $n_{\rm eff}$.

\subsection{Repeat Observations and Statistical Redshift Uncertainty}
\label{subsec:repeats}
The spectroscopic measurements of redshift have associated uncertainties (i.e., the redshift uncertainty) due to factors like the spectral line width, observing conditions, and different astrophysical effects. The impact of the redshift uncertainty is equivalent to adding stochasticity to the peculiar velocity of the observed object, and thus will bias the measurement of anisotropic clustering and velocity bias \citep{Guo2015b,  zerr2018, mine}. The redshift uncertainty can be quantified by repeat observations statistically and via its influence on the clustering using our SHAM method (see Section~\ref{subsec:sham}). The results of those two estimators should be consistent. 

Objects observed repetitively exist in all stages of Survey Validation. However, the ones from the One Percent Survey are biased towards faint objects by design (see Section~\ref{subsec:EDR}). Therefore, we used data from the early stage of Survey Validation to obtain an unbiased estimate of the redshift uncertainty. The redshift difference, $\Delta z$, is calculated for all pairs of repeated spectra for each object and then converted to radial velocity using $\Delta v = c\Delta z/(1+z)$, where $c$ is the speed of light and $z$ is the mean redshift of the pairs. $\Delta v$ measurements larger than the redshift failure threshold ($1000\kms$ for LRGs and ELGs, $3000\kms$ for QSOs) are then removed. 

The histograms of the redshift difference of ELGs, LRGs, and QSOs are presented in \reffig{fig:deltav measurement} in black dots. The error bars of those histograms are calculated using the delete-one jackknife method. Our fitting range is around $[-200,200]\kms$ for LRGs, $[-150,150]\kms$ for ELGs, and $[-1600,1600]\kms$ for QSOs except for $[-500,500]\kms$ for QSOs at $0.8<z<1.1$. The title of each subplot in \reffig{fig:deltav measurement} shows the percentage of $\Delta v$ measurements that are beyond the fitting range. For SDSS BOSS/eBOSS LRGs, the redshift difference in all redshift ranges can be well fitted by Gaussian functions \citep{ross_completed_2020,lyke20a,mine}. For DESI, all tracers show a preference for Lorentzian distributions in general (solid red lines in \reffig{fig:deltav measurement}) as
\begin{equation}
    \mathcal{L}(p,w_{\Delta v}) = \frac{A}{1+((x-p)/w_{\Delta v})^2}.
    \label{eq:lorentzian}
\end{equation}
In \refeq{eq:lorentzian}, $A$ is a normalization factor, $p$ is the location of the peak value on the x-axis, and $2w_{\Delta v}$ is the full-width-half-maximum of the Lorentzian distribution. In addition, we also try to describe $\Delta v$ histograms of LRGs with Gaussian profiles $\mathcal{N}(\mu,\sigma_{\Delta v})$ (solid blue lines in \reffig{fig:deltav measurement}). We will discuss which profile to use for SHAM in Section~\ref{subsec:vsmear and deltav}. As $p$ and $\mu$ are well consistent with 0, we only present the best fitting Lorentzian $w_{\Delta v}$ and Gaussian $\sigma_{\Delta v}$ on the labels of \reffig{fig:deltav measurement}. We also calculate the standard deviation of the redshift difference $\hat{\sigma}_{\Delta v}$.

In \reffig{fig:deltav measurement}, we observe a much smaller redshift uncertainty for the ELGs than that of the LRGs and QSOs. The maximum  $w_{\Delta v}$ of ELGs is $13.4\pm 0.1\kms$, while the minimum  $w_{\Delta v}$ of LRG and QSOs is $34.6\pm 0.9$ and $30.2\pm0.2\kms$ respectively. This can be attributed to the narrow \oii{} emission for ELG redshift determination, compared with absorption lines of LRGs and broad emissions of QSOs. Additionally, galaxy samples (LRGs and ELGs) show increasing uncertainty with redshift. This is because galaxies are fainter at higher redshift, and spectral lines for redshift determination have a decreasing signal-to-noise ratio and larger uncertainty. But for QSO this is not the case, as QSOs at higher redshifts are not necessarily fainter. Another reason for the non-monotonic QSO $w_{\Delta v}$ trend is that the measurement made by repeat observation is no longer reliable at $z\gtrsim 1.5$. We will explain this in detail in Section~\ref{subsec:vsmear and deltav}. 


\subsection{\texorpdfstring{$N$}{N}-body Simulation: UNIT}
\label{subsec:sim}
We apply our SHAM on Universe $N$-body simulations for the Investigation of Theoretical models from galaxy surveys\footnote{\url{http://www.unitsims.org/}} \citep[UNIT;][]{UNIT} to generate model galaxies in cubic boxes. Planck cosmology \citep{planck2016} is employed in the UNIT simulations and our SHAM implementation: $\Omega_{\rm m} = 0.3089,~h \equiv H_0/100\,{\rm km}\,{\rm s}^{-1}\,{\rm Mpc}^{-1} = 0.6774,~ n_s = 0.9667,~ \sigma_8 = 0.8147$. In each 1$\,h^{-3}\,{\rm Gpc}^{3}$ UNIT simulation box, there are $4096^3$ particles with the mass resolution of $1.2\times10^9\,\Msun{}$. 

We use UNIT halo catalogues with subhaloes identified by the \textsc{rockstar} \citep{Behroozi2013a} halo finder that provides properties at the current snapshot, such as positions, peculiar velocities, virial mass $M_{\rm vir}$, and the maximum circular velocity $V_{\rm max}$. We regard $M_{\rm vir}$ of haloes with more than 50 dark matter particles to be reliable, i.e., $M_{\rm vir}^{\rm good}>6\times10^{10}\Msun{}$. The merger/stripping histories of haloes and subhaloes are provided by \textsc{consistent trees} \citep{Behroozi2013b}. They are used to determine their peak maximum circular velocity throughout the accretion history, i.e., $V_{\rm peak}$, which is the proxy of halo mass in our SHAM study.

UNIT simulations are created using the fixed-amplitude method implemented in pairs of simulation boxes to suppress the cosmic variance \citep{Angulo2016, UNIT}. The effective volume of UNIT simulations is much larger than those of DESI EDR tracers as shown in Table~\ref{tab:3param result}. So we can take just one simulation box in each snapshot and ignore the influence of the UNIT cosmic variance on our SHAM fitting.

UNIT includes 128 snapshots of simulations from redshift 99 to 0 and we employ 14 of them with their redshift presented in the fourth column of Table~\ref{tab:3param result}. We select the UNIT snapshot whose redshift is the closest to the $z_{\rm eff}$ (\refeq{eq:zeff}) of the corresponding DESI sample among all the snapshots.

\section{Method}
\label{sec:method}  
\subsection{Galaxy Clustering}
\label{subsec:2pcf}
The two-point correlation function (2PCF) measures the excess probability of finding a galaxy pair compared to a random distribution in a given volume. For observations, we use the Landy--Szalay estimator \citep[LS;][]{Landy1993} which minimises the variances of the measurements for an irregular geometry: 
\begin{equation}
\label{LS estimator}
   \xi_{\rm LS}  = \frac{\rm DD-2DR+RR}{\rm RR}, 
\end{equation}
where the data--data (DD), data--random (DR), and random--random (RR) pair counts are normalized by their corresponding total number of pairs. $\xi$ and the pair counts can be calculated as a function of the pair separation $s$ and $\mu$ which is the cosine of the angle between the line connecting the galaxy pairs and the line-of-sight. $w_{\rm FKP}$ is applied to every individual galaxy in the data and random catalogue. PIP weights are applied to DD pair counts, 
and ANG weights are implemented to both the DD and DR pairs. 

The SHAM galaxies are populated in periodic boxes based on halo catalogues from the UNIT $N$-body simulation, so we use the Peebles--Hauser estimator \citep[PH;][]{Peebles1974} to obtain their 2PCF as follows: 
\begin{equation}
    \label{PH estimator}
    \xi_{\rm PH} = \frac{\rm DD}{\rm RR}-1,
\end{equation}
Unlike observation that requires random catalogues to calculate RR pairs, we use the following expression to calculate them analytically in the simulation box:
\begin{equation}
\label{RR(s,mu)}
    \rm RR = \frac{4\uppi}{3}\frac{\textit{s}_\text{max}^3-\textit{s}_\text{min}^3}{\textit{V}_\text{box}} \frac{1}{\textit{N}_{\mu}},
\end{equation}
where $s_\text{max}$ and $s_\text{min}$ are the boundaries of the separation bins, $V_\text{box}=1\,h^{-3}\,{\rm Gpc}^{3}$ is the volume of the UNIT simulation box, and $N_{\mu}=200$ is the number of $\mu$ bins. 

By weighting the 2D $\xi(s,\mu)$ with Legendre polynomials $P_\ell(\mu)$, we obtain the 1D $\xi$ multipoles as 
\begin{equation}
    \label{multipoles}
    \xi_\ell (s) = \frac{2\ell+1}{2}\int_{-1}^1 \xi(s,\mu) P_\ell(\mu) {\rm d}\mu.
\end{equation}
We fit our SHAM to observations based on the monopole and quadrupole, i.e., $\ell=0,2$. We use 10 logarithmic $s$ bins in $(5,30)\hmpc$ and 200 $\mu$ bins in $(-1,1)$.

The projected 2PCF is calculated for cross-checking the clustering of the best-fitting SHAM galaxies. This is calculated as
\begin{equation}
    w_p(r_p) = \int_{-\pi_{\rm max}}^{\pi_{\rm max}}\xi(r_p,\pi){\rm d}\pi,
\end{equation}
where $\pi_{\rm max}=30\hmpc$ to avoid the contamination of the systematics on larger scales as shown in \citetalias{mine}. \textsc{pycorr} and \textsc{Corrfunc} Python packages \citep{10.1007/978-981-13-7729-7_1,2020MNRAS.491.3022S} are used to calculate $\xi_\ell(s)$ and $w_p$.

In observations, 2PCFs are calculated in redshift space. So the position of our mock galaxies produced by SHAM should take into account the redshift-space distortion \citep[RSD;][]{Kaiser1987} using:
\begin{equation}
    Z_{\rm redshift}=Z_{\rm real}+\frac{v_{\rm pec,Z}(1+z)}{H(z)}, 
    \label{real-redshift}
\end{equation} 
where $Z$ is the coordinate in the $Z$-axis which is the line of sight, and its subscripts `redshift' and `real' illustrate that the coordinate is in the redshift space or in the real space. $v_{\rm pec,Z}$ is the proper peculiar velocity of SHAM galaxies along the $Z$-axis, and $z$ is the redshift of the simulation snapshot. As the cosmic variance of UNIT simulations is small, we can safely ignore the variations in quadrupoles for different line-of-sights \citep{xi2losbias_2021}.

\subsection{SHAM Implementation}
\label{subsec:sham}
SubHalo Abundance Matching (SHAM) is an empirical method to construct a realistic, monotonic galaxy--halo mass relation based on $N$-body simulations. In its simplest form, a SHAM has a single free parameter $\sigma$ relating the masses of galaxies and haloes and can successfully reproduce the observed clustering \citep[e.g.,][]{tasitsiomi_modeling_2004,Behroozi2010}. As observations provide the clustering of multiple tracers with higher and higher accuracy, this prototype should also be improved. We thus introduce the massive (sub)halo incompleteness $V_{\rm ceil}$, the redshift uncertainty $v_{\rm smear}$, and a free satellite fraction $f_{\rm sat}$ in the SHAM implementation besides the galaxy--halo mass scatter $\sigma$. Their impact on the 2PCF $\xi_{\ell}(s)$ and projected 2PCF $w_p(r_p)$ are presented in Appendix~\ref{appendix:LOWZ SHAM}. 

In our study, all (sub)haloes in the simulation have their $V_{\rm peak}$ multiplied by an asymmetric Gaussian as
\begin{equation}
\label{eq:Vpeak_scat}
    V_{\rm scat} = V_{\rm peak} \times
    \begin{cases}
        1+\mathcal{N}(0,\sigma), &  \mathcal{N}(0,\sigma)>0; \\
        \exp(\mathcal{N}(0,\sigma)), & \mathcal{N}(0,\sigma)<0,
    \end{cases}
\end{equation}
to avoid negative $V_{\rm scat}$. 

Then those (sub)haloes are sorted in descending order of $V_{\rm scat}$ and the first $V_{\rm ceil}N_{\rm UNIT}/100$ ones are removed. $N_{\rm UNIT}$ is the total number of haloes and subhaloes in this UNIT simulation. It means the most massive (sub)haloes will not be assigned with a galaxy/QSO in its centre. $V_{\rm ceil}$ is introduced for target selections that possibly remove some of the most massive LRGs, resulting in incompleteness in the host (sub)halo mass. ELGs are mostly star-forming galaxies and thus are not expected to be complete in stellar mass and thus (sub)halo mass \citep[e.g][]{violeta_number_density, boryana2021}. This is because the hot and dense centre of massive (sub)haloes is an environment that depletes the cold gas and thus stops star formation \citep[e.g.][]{kauffmann_environmental_2004,Dekel2006,peng2010}. QSOs are bright active galactic nuclei (AGN), i.e., their super-massive black holes actively accrete cold gas via discs \citep[e.g.,][]{QSOproperty2013}. In the semi-analytical models (SAM), the formation of AGNs with $L_{\rm bol}\gtrsim10^{45.1}\,\rm erg\,s^{-1}$ only happens at haloes with $M_{\rm vir}\lesssim 10^{13}\,\Msun{}$ during starbursts \citep{griffin2019}. \citet{QSOincomplete2017} attribute the absence of QSOs in the overdense regions (i.e., massive haloes) at $z\sim2$--3 to the lack of wet mergers which leads to the QSO activity. In hydrodynamical simulations, \citet{QSOincomplete_hydro} also find that AGNs exit their high-accretion phase (i.e., the QSO phase) in the most massive galaxy at $z\sim2$. The absence of QSOs in those massive quenched galaxies means their absence in the most massive haloes. 
So LRGs, ELGs and QSOs all require the $V_{\rm ceil}$ truncation, which still allows (sub)haloes with large $V_{\rm peak}$ with the help of $\sigma$. We need to point out that the actual format of the massive halo incompleteness should not depend solely on $V_{\rm peak}$. This $V_{\rm ceil}$ truncation is chosen as it is the simplest implementation for the incompleteness and it enables a good description of the observed 2-point clustering (See Section~\ref{subsec:clustering}). $v_{\rm smear}$-SHAM and $f_{\rm sat}$-SHAM algorithms then deviate after this step. 

For $v_{\rm smear}$-SHAM, we populate a central/satellite galaxy in the centre of each halo/subhalo in the $V_{\rm ceil}$-truncated catalogue from the most massive ones to the least ones until we get the expected number of SHAM galaxies 
\begin{equation}
    \label{eq:Ngal}
    N_{\rm gal}=n_{\rm eff}V_{\rm box},
\end{equation}
where $V_{\rm box}=1 \hgpc$ is the box size of the UNIT $N$-body simulation. $n_{\rm eff}$ is the effective number density of the observed sample obtained using \refeq{eq:neff} and the values for each galaxy sample are presented in Table~\ref{tab:3param result}.
The proper peculiar velocity of the host (sub)haloes $\boldsymbol{v}_{\rm pec}^h$ is also assigned to their galaxies. The velocity of the galaxy along the line of sight $v_{\rm pec,Z}^g$ is then blurred by $v_{\rm smear}$ to mimic the effect of the redshift uncertainty as
\begin{equation}
    v_{\rm pec,Z}^g=v_{\rm pec,Z}^h+    
    \begin{cases}
        \mathcal{N}(0,v_{\rm smear,G}), &  \text{Gaussian profile};\\
        \mathcal{L}(0,v_{\rm smear,L}), & \text{truncated Lorentzian profile},
    \end{cases}
\end{equation}
where $v_{\rm pec,Z}^h$ is the component of $\boldsymbol{v}_{\rm pec}^h$ on the $Z$-axis, $\mathcal{N}(0,v_{\rm smear,G})$ and $\mathcal{L}(0,v_{\rm smear,L})$ are a random number sampled by a Gaussian profile or a Lorentzian profile, respectively (as discussed in Section~\ref{subsec:repeats}). As the standard Lorentzian profile is heavy-tailed and subexponential, we remove $\mathcal{L}(0,v_{\rm smear})$ larger than $400\,\kms$ for LRGs and $2000\kms$ for QSOs. We do not use the $v_{\rm smear}$ parameter in $f_{\rm sat}$-SHAM as explained in Section~\ref{subsec:vsmear and deltav}. 

In $f_{\rm sat}$-SHAM, we further separate haloes and subhaloes from the $V_{\rm ceil}$-truncated catalogue. Only the first $f_{\rm sat}N_{\rm gal}/100$ subhaloes are kept as hosts of ELG satellites and the first $(1-f_{\rm sat})N_{\rm gal}/100$ haloes are for central ELGs. Then we assign the centre position and the proper peculiar velocity of those selected halo/subhalo to their central/satellite galaxies. 

Note that in $v_{\rm smear}$-SHAM, the satellite fraction $f_{\rm sat}$ is defined as the percentage of subhaloes in the list of (sub)haloes selected by SHAM, that is, 
\begin{equation}
\label{eq:fsat}
    f_{\rm sat}\equiv\frac{N_\text{sat}}{N_\text{gal}}=\frac{N_{\rm sub,SHAM}}{N_{\rm sub, SHAM}+N_{\rm halo,SHAM}}  
\end{equation}
where $N_{\rm sat}$ is the number of satellite galaxies in the SHAM galaxy catalogue, $N_{\rm sub,SHAM}=N_{\rm sat}$ is the number of subhaloes selected by SHAM, and $N_{\rm halo,SHAM}$ is the number of haloes selected by SHAM. Note that $f_{\rm sat}$ is different from the percentage of subhaloes in the UNIT simulations:
\begin{equation}
\label{eq:fsub}
    f_{\rm sub}\equiv\frac{N_\text{sub}}{N_{\rm UNIT}}=\frac{N_{\rm sub}}{N_{\rm sub}+N_{\rm halo}} 
\end{equation}
where $N_\text{sub}$ is the total number of subhaloes in the UNIT simulation and $N_{\rm halo}$ is the total number of haloes there. 

Finally, we calculate the clustering of model galaxies in redshift space produced by $v_{\rm smear}$-SHAM (LRGs, QSOs) or $f_{\rm sat}$-SHAM (ELGs) and compare it with observations, trying to find the best-fitting parameters. As shown in Appendix~\ref{appendix:LOWZ SHAM}, $\sigma$, $V_{\rm ceil}$, $v_{\rm smear}$ and $f_{\rm sat}$ are the primary factors that affect the spatial distribution of DESI dark matter tracers at 5-30$\hmpc$. Given the well-reproduced clustering (see Section~\ref{subsec:clustering}), we do not explore additional effects such as the assembly bias, which can not be well constrained by our DESI sample due to its low number density  \citep{2021MNRAS.504.5205C,Yuan2023,Rocher2023}. However, to describe the galaxy-halo connection of dense tracers like the Bright Galaxy Sample \citep[BGS;][]{Pearl2023}, and the cross-correlation between different tracers (\citet{Gao2023}, Yuan et al. in prep), assembly bias will play a role. In addition to the 3-parameter SHAM, we further discuss the performance of the complete 4-parameter SHAM \{$\sigma,V_{\rm ceil},v_{\rm smear}, f_{\rm sat}$\} in Appendix~\ref{appendix:LOWZ SHAM}. 
\subsection{SHAM Constraints}
\label{subsec:fitting}
We try to find the best-fitting SHAM parameters using a Monte-Carlo sampler \textsc{Multinest}\footnote{\url{https://github.com/farhanferoz/MultiNest}} \citep{Feroz2008,2009MNRAS.398.1601F,2019OJAp....2E..10F} assuming a Gaussian likelihood $\mathcal{L}(\Theta)$ for our parameter constraint
\begin{equation}
\label{likelihood}
    \mathcal{L}(\Theta) \propto \rm e^{-\frac{\chi^2(\Theta)}{2}}.
\end{equation}
The $\chi^2$ values are obtained as
\begin{equation}
\chi^2 (\Theta) =  (\boldsymbol{\xi}_{\rm data}-\boldsymbol{\xi}_{\rm model}(\Theta))^T\textbf{C}^{-1}(\boldsymbol{\xi}_{\rm data}-\boldsymbol{\xi}_{\rm model}(\Theta)),
\end{equation}
where $\Theta = \{\sigma,V_{\rm ceil},v_{\rm smear}\}$ for LRG and QSO samples and $\Theta = \{\sigma,V_{\rm ceil},f_{\rm sat}\}$ for ELG samples. $\boldsymbol{\xi} = (\xi_0,\xi_2)$ denotes the vector composed of the 2PCF monopole and quadrupole. The subscripts `data' and `model' of $\boldsymbol{\xi}$ represent measurements from the observational data and SHAM mocks, respectively. 
\textbf{C} is the unbiased covariance matrix that should include the variances of $\boldsymbol{\xi}_{\rm data}$ and $\boldsymbol{\xi}_{\rm model}$. The variances of $\boldsymbol{\xi}_{\rm model}$ can be further decomposed into the cosmic variance of UNIT simulation and the statistical variance due to the random processes included (Section~\ref{subsec:sham}). The variance of UNIT is considered to be negligible (Section~\ref{subsec:sim}). $\boldsymbol{\xi}_{\rm model}$ is obtained by averaging the 2PCFs of 32 SHAM galaxy realizations generated using the same $\Theta$ with different random seeds. 
Because the statistical variance of 32 realizations is less than 5 per cent of the observational errors in general, increasing the number would increase the computing cost without much gain in the reliability of the parameter constraint. So \textbf{C} can be estimated as the variance of $\boldsymbol{\xi}_{\rm data}$ via \citep{Hartlap2007}: 
\begin{equation}
    \textbf{C}^{-1} = \textbf{C}^{-1}_{\rm s} \frac{N_{\rm mocks}-N_{\rm bins}-2}{N_{\rm mock}-1},
\end{equation}
where $N_{\rm bins}=20$ (Section~\ref{subsec:2pcf}) is the length of $\boldsymbol{\xi}_{\rm data}$, i.e., the total number of bins of the monopole and the quadrupole used in the SHAM fitting. $\textbf{C}_{\rm s}$ is the jackknife covariance matrix, and is calculated using \textsc{pycorr}\footnote{\url{https://github.com/cosmodesi/pycorr}} with $N_{\rm mock}=128$ jackknife subsamples of the observational data.
$\textbf{C}_{\rm s}$ is thus expressed as
\begin{equation}
\label{eq:jk cov}
   \textbf{C}_{\text{s},ij} = \frac{1}{N_{\rm mock}-1}\sum^{N_{\rm mock}}_{k=1} [\boldsymbol{\xi}_{i}^{(k)}-\overline{\boldsymbol{\xi}}_i][\boldsymbol{\xi}_{j}^{(k)}-\overline{\boldsymbol{\xi}}_j],
\end{equation}
where $\boldsymbol{\xi}^{(k)}$ is the correlation function measured from the data with the $k_{\rm th}$ jackknife subsample removed, and 
\begin{equation}
\overline{\boldsymbol{\xi}}_i = \frac{1}{N_{\rm mock}}\sum^{N_{\rm mock}}_{k=1}\boldsymbol{\xi}_i^{(k)}
\end{equation}
is the mean 2PCF of all jackknife subsamples. The errors for the data vector are the square root of the diagonal terms of $\textbf{C}$.

We employ \textsc{Multinest}, an efficient nested sampling technique, to constrain $\Theta$. We keep the default convergence criteria which is a tolerance of 0.5 and use 200 particles for the sampling. Using a smaller tolerance or more particles takes more computing time but provides a similar posterior. The prior range for the SHAM fitting is listed in Table~\ref{tab:prior}.
\begin{table}
    \centering
    \begin{tabular}[c]{c|c|c|c|c}
    \hline 
    \hline 
    {tracer}&{$\sigma$}&{$v_{\rm smear,L}$}&{$V_{\rm ceil}$}&{$f_{\rm sat}$}\\
    {}&{}&{$(\kms)$}&{$(\%)$}&{$(\%)$}\\
    \hline
    {LRG}&{[0,1]}&{[0,200]}&{[0,0.15]}&{/ } \\
    \hline
    {QSO}&{[0,2]}&{[0,1600]}&{[0,2]}&{/}\\
    \hline
    {ELG}&{[0,1]}&{/}&{[0,20]}&{[0,30]}\\
    \hline    
    \end{tabular}
    \caption{The priors of $v_{\rm smear}$-SHAM for LRGs and QSOs, and of $f_{\rm sat}$-SHAM for ELGs. Priors of $v_{\rm smear}$-SHAM with a Gaussian profile $v_{\rm smear,G}$ are the same with those with $v_{\rm smear,L}$.}
    \label{tab:prior}
\end{table}
 The best-fitting parameters, which are the medians of the 16th and 84th percentiles (the 1-$\sigma$ confidence limit) of the marginalized posterior distributions of individual parameters (Appendix~\ref{appendix:Posterior Contours}), their 1-$\sigma$ confidence limits, and the minimum $\chi^2$ are provided by \textsc{PyMultinest}\footnote{\url{https://github.com/JohannesBuchner/PyMultiNest}} \citep{pymultinest}. All the derived quantities, i.e., $f_{\rm sat}$ (in $v_{\rm smear}$-SHAM), halo occupation distribution (HOD), the probability of a (sub)halo to host a central (satellite) galaxy (PDF), the mean halo mass $\langle M_{\rm vir}\rangle$ and the mean $V_{\rm peak}$ $\langle V_{\rm peak} \rangle$ are computed using the nested sampling chain.

\begin{table*} 
{
\begin{tabular}[c]{|c|c|c|c|c|c|c|c|c|c|c|}
\hline 
\hline 
{tracer}&{redshift} & {$z_{\rm eff}$} & {$z_{\rm UNIT}$} &{ $V_{\rm eff}$}& {$n_{\rm eff}\times10^4$} & {$\sigma$} &  {$v_{\rm smear,L}$} & {$V_{\rm ceil}$}& {$f_{\rm sat}$} &{$\chi^2$/dof}\\ 
{type}  &{range}    & {} & {} &{ ($\hgpc$)}& {$({\rm Mpc}^{-3}\,h^3)$} & {}  & {$(\kms)$} &{$(\%)$}&{$(\%)$}& {} \\ 
\hline 
\hline 
{LRG}&{$0.4<z<1.1$}&{0.8138} & {0.8188} &{0.150} & {5.50}&{$0.27^{+0.09}_{-0.11}$}&{$40^{+9}_{-9}$}&{$0.02^{+0.01}_{-0.01}$}&{$13.9^{+0.4}_{-0.4}$}&{24/17} \\ 
\hline 
{ELG}&{$0.8<z<1.6$}&{1.2020} & {1.2200} &{0.204} & {7.26}&{$0.28^{+0.24}_{-0.18}$}&{/}&{$3.62^{+0.88}_{-1.10}$}&{$3.4^{+1.9}_{-1.6}$}&{22/17} \\ 
\hline 
{QSO}&{$0.8<z<3.5$}&{1.7408} & {1.7710} &{0.024} & {0.24}&{$0.38^{+0.31}_{-0.24}$}&{$215^{+24}_{-31}$}&{$0.23^{+0.11}_{-0.09}$}&{$12.3^{+0.5}_{-0.5}$}&{9/17} \\ 
\hline 
\hline
{LRG}&{$0.4<z<0.6$}&{0.5126} & {0.5232} &{0.032} & {6.16}&{$0.16^{+0.16}_{-0.11}$}&{$27^{+17}_{-15}$}&{$0.04^{+0.02}_{-0.02}$}&{$15.2^{+0.5}_{-0.5}$}&{15/17} \\ 
\hline 
{LRG}&{$0.6<z<0.8$}&{0.7067} & {0.7018} &{0.052} & {6.87}&{$0.25^{+0.07}_{-0.10}$}&{$19^{+12}_{-9}$}&{$0.01^{+0.01}_{-0.01}$}&{$14.5^{+0.4}_{-0.4}$}&{23/17} \\ 
\hline 
{LRG}&{$0.8<z<1.1$}&{0.9423} & {0.9436} &{0.076} & {4.36}&{$0.28^{+0.15}_{-0.14}$}&{$30^{+14}_{-12}$}&{$0.04^{+0.03}_{-0.02}$}&{$13.2^{+0.5}_{-0.5}$}&{29/17} \\ 
\hline 
{ELG}&{$0.8<z<1.1$}&{0.9565} & {0.9436} &{0.088} & {10.47}&{$0.31^{+0.43}_{-0.21}$}&{/}&{$6.68^{+2.17}_{-2.69}$}&{$5.5^{+2.5}_{-2.2}$}&{18/17} \\ 
\hline 
{ELG}&{$1.1<z<1.6$}&{1.3397} & {1.3210} &{0.121} & {5.13}&{$0.27^{+0.31}_{-0.17}$}&{/}&{$3.07^{+0.80}_{-1.09}$}&{$4.2^{+2.4}_{-2.1}$}&{22/17} \\ 
\hline 
{QSO}&{$0.8<z<1.1$}&{0.9658} & {0.9436} &{0.003} & {0.29}&{$0.52^{+0.49}_{-0.30}$}&{$101^{+87}_{-59}$}&{$0.67^{+0.52}_{-0.23}$}&{$20.1^{+1.2}_{-1.2}$}&{17/17} \\ 
\hline 
{QSO}&{$1.1<z<1.6$}&{1.3665} & {1.3720} &{0.009} & {0.36}&{$0.32^{+0.49}_{-0.21}$}&{$78^{+35}_{-26}$}&{$0.45^{+0.21}_{-0.21}$}&{$15.9^{+0.8}_{-0.8}$}&{11/17} \\ 
\hline 
{QSO}&{$1.6<z<2.1$}&{1.8320} & {1.8330} &{0.008} & {0.31}&{$0.25^{+0.44}_{-0.18}$}&{$273^{+151}_{-61}$}&{$0.26^{+0.09}_{-0.15}$}&{$11.5^{+0.8}_{-0.8}$}&{13/17} \\ 
\hline 
{QSO}&{$2.1<z<3.5$}&{2.4561} & {2.4580} &{0.004} & {0.13}&{$0.29^{+0.45}_{-0.18}$}&{$542^{+75}_{-100}$}&{$0.14^{+0.07}_{-0.09}$}&{$8.1^{+0.7}_{-0.7}$}&{12/17} \\ 
\hline 
\hline 
\end{tabular} } 
\caption{The information for observation and its best-fitting SHAM results of $v_{\rm smear}$-SHAM with Lorentzian redshift uncertainty profile $v_{\rm smear,L}$ and $f_{\rm sat}$-SHAM. The columns are: 1) observed tracer type, 2) redshift range, 3) effective redshift calculated using \refeq{eq:zeff}, 4) redshift of the UNIT simulation snapshot for the SHAM fitting that is close to $z_{\rm eff}$, 5) the effective volume $V_{\rm eff}$ of the observed tracer at the corresponding redshift range obtained with \refeq{eq:Veff}, 6) the effective number density $n_{\rm eff}$ calculated with \refeq{eq:neff} multiplied by $10^4$, the best-fitting parameters, i.e., 7) $\sigma$, 8) the redshift uncertainty $v_{\rm smear,L}$, 9) the massive-(sub)halo incompleteness $V_{\rm ceil}$, 10) the satellite fraction $f_{\rm sat}$ and 11) the minimum $\chi^2$ divided by the degree of freedom. $v_{\rm smear}$ of ELG samples are asserted to 0 and the satellite fraction of LRGs and QSOs is a derived parameter from the nested sampling chain. 
} 
\label{tab:3param result} 
\end{table*} 

\begin{table} 
\scalebox{0.88}
{
\begin{tabular}[c]{|c|c|c|c|c|c|c|}
\hline 
\hline 
{redshift} & {$\sigma$} &  {$v_{\rm smear,G}$} & {$V_{\rm ceil}$}& {$f_{\rm sat}$} &{$\chi^2$/dof}\\ 
{range}    & {}  & {$(\kms)$} &{$(\%)$}&{$(\%)$}& {} \\ 
\hline 
\hline 
{$0.4<z<1.1$}&{$0.21^{+0.14}_{-0.10}$}&{$95^{+12}_{-14}$}&{$0.03^{+0.02}_{-0.02}$}&{$14.7^{+0.3}_{-0.3}$}&{24/17} \\ 
\hline 
\hline
{$0.4<z<0.6$}&{$0.14^{+0.11}_{-0.08}$}&{$67^{+28}_{-30}$}&{$0.04^{+0.02}_{-0.02}$}&{$15.2^{+0.4}_{-0.4}$}&{15/17} \\ 
\hline 
{$0.6<z<0.8$}&{$0.31^{+0.04}_{-0.06}$}&{$42^{+16}_{-22}$}&{$0.00^{+0.01}_{-0.00}$}&{$14.7^{+0.3}_{-0.3}$}&{23/17} \\ 
\hline 
{$0.8<z<1.1$}&{$0.28^{+0.19}_{-0.17}$}&{$84^{+17}_{-18}$}&{$0.04^{+0.03}_{-0.03}$}&{$14.7^{+0.3}_{-0.3}$}&{30/17} \\ 
\hline 

\end{tabular} } 
\caption{The best-fitting results of the $v_{\rm smear}$-SHAM fitting for LRGs with Gaussian $v_{\rm smear,G}$.} 
\label{tab:gaussian result} 
\end{table} 

\section{Results}
\label{sec:results}
We present results of $v_{\rm smear}$-SHAM for LRGs and QSOs, $f_{\rm sat}$-SHAM for ELGs for the DESI One Percent Survey in this section. The best-fitting 2PCF, features of the best-fitting $\sigma$, $V_{\rm ceil}$, $v_{\rm smear}$ and $f_{\rm sat}$ for different tracers are discussed respectively. We also check the consistency between the HOD measured from our best-fitting SHAM with those from HOD studies using the same data. 
\begin{figure}
\includegraphics[width=\columnwidth]{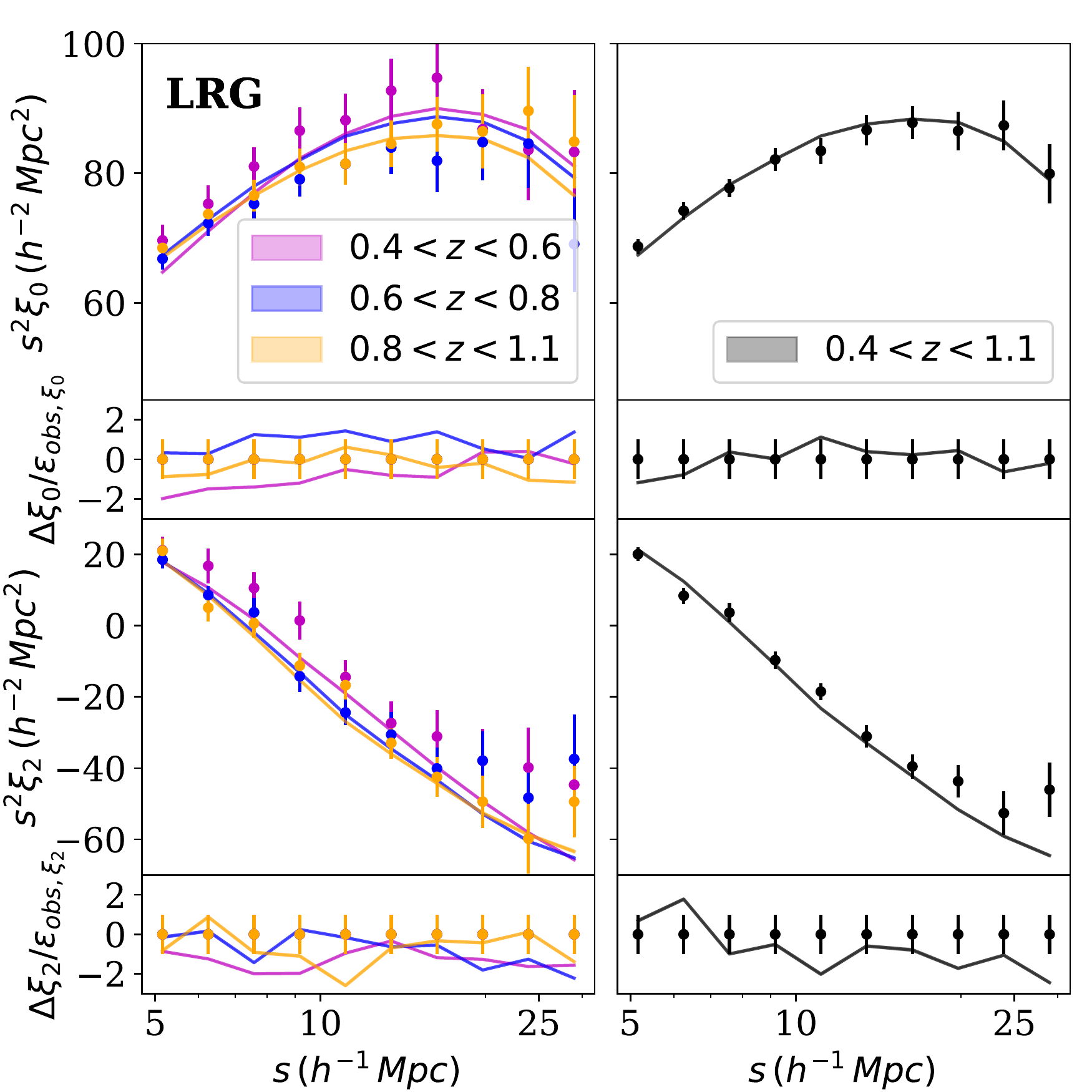}
    \caption{The clustering of observed LRGs (filled circles with error bars) compared with that of the best-fitting SHAM model galaxies with its statistical uncertainty (solid lines with shades). Monopoles and their residuals normalised by the observed errors $\epsilon_{\rm obs}$ are presented in the first and the second rows. The third and fourth rows present those for quadrupoles. Each colour shows a different redshift range, as indicated in the legend. The first column shows results for LRGs in different redshift bins and the second for the total sample. 
    The error bars of data are obtained from 128 jackknife samples, and the statistical uncertainty of SHAM galaxies indicated in the width of the shades is the standard deviation of its 32 realizations divided by $\sqrt{32}$. The uncertainty of best-fitting LRG SHAM galaxies is too small to be seen. Our SHAM provides good fit to the observed clustering at 5--30$\hmpc$.}
    \label{fig:lrg 2pcf}
\end{figure}
\begin{figure}
	\includegraphics[width=\columnwidth]{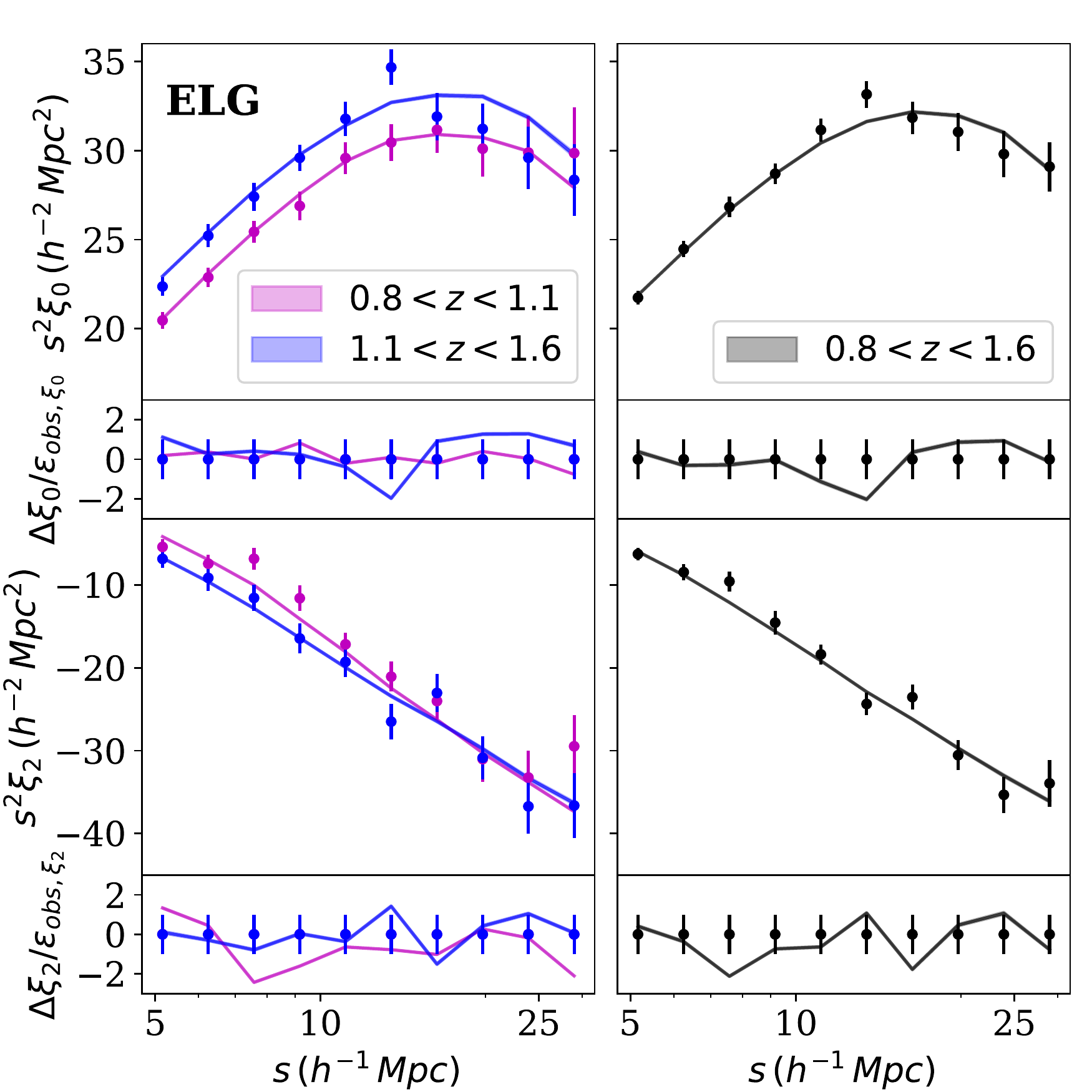}
    \caption{Same as \reffig{fig:lrg 2pcf} but for ELGs.}
    \label{fig:elg 2pcf}
\end{figure}
\begin{figure}
	\includegraphics[width=\columnwidth]{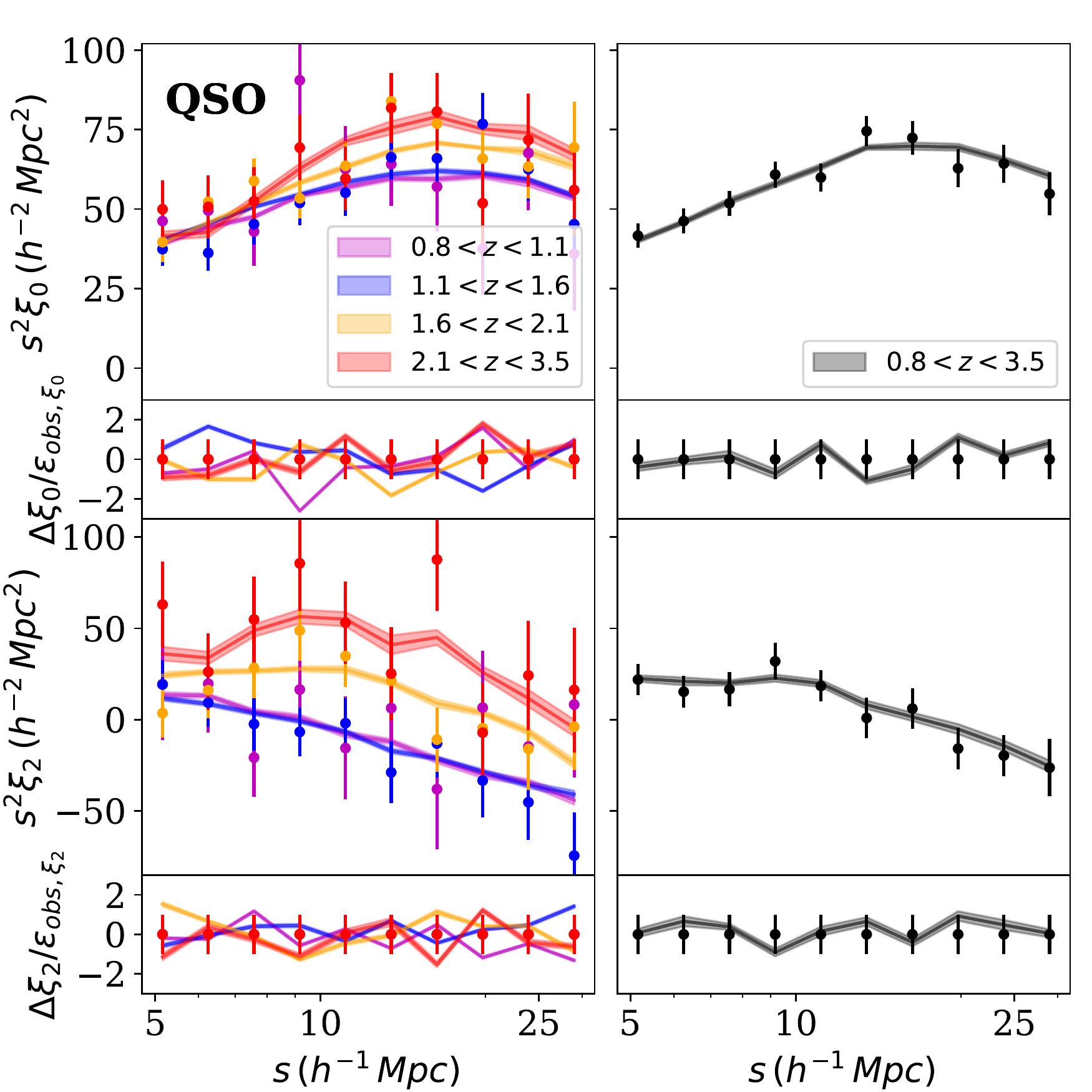}
    \caption{Same as \reffig{fig:lrg 2pcf} but for QSOs.}
    \label{fig:qso 2pcf}
\end{figure}
\begin{figure}
\includegraphics[scale=0.6]{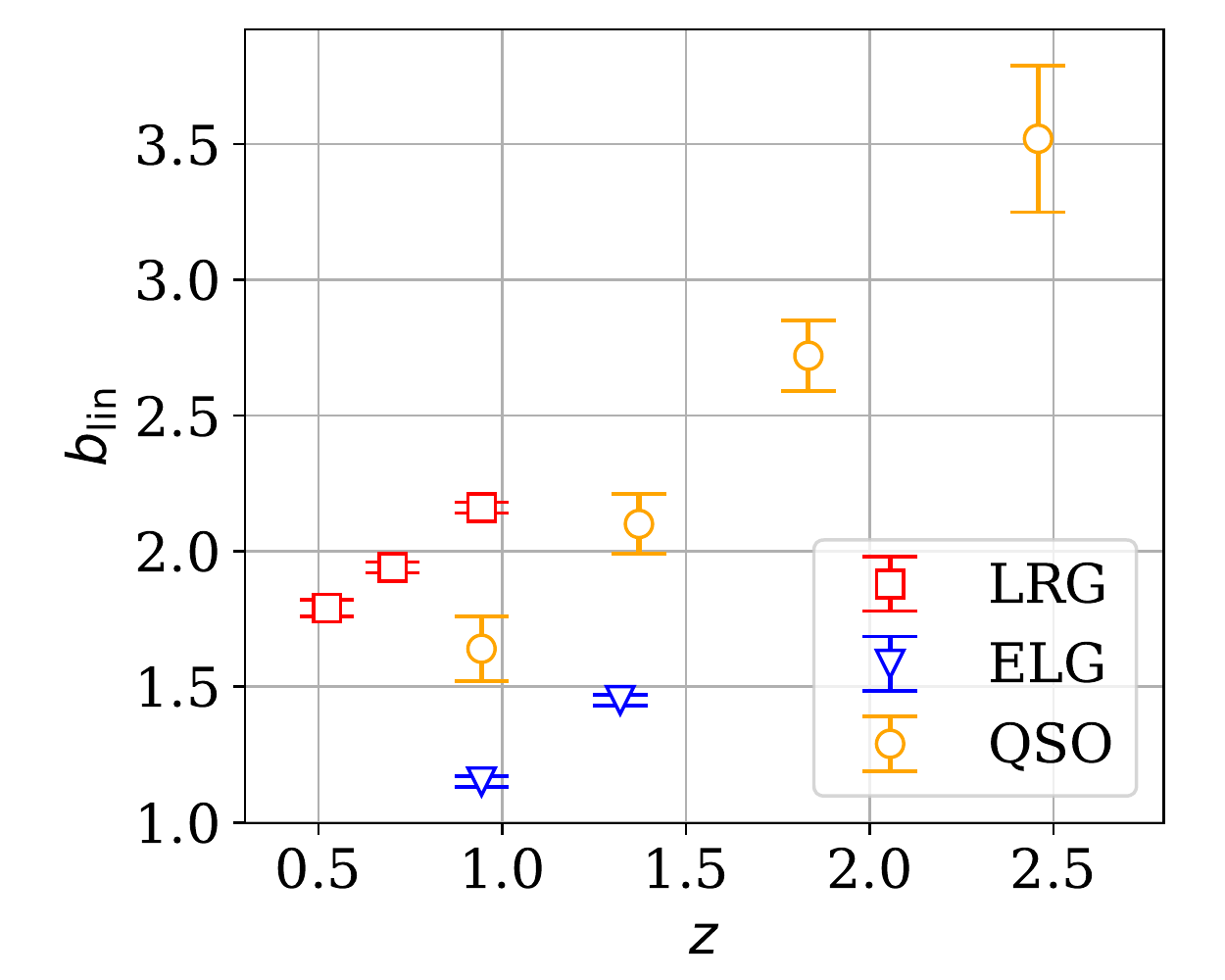}
    \caption{The linear bias of the best-fitting model galaxies and quasars from SHAM with error bars. They are calculated via the power spectrum at $k<0.05\,h\,\rm Mpc^{-1}$. For each type of tracer, the linear bias increases with the redshift. }
    \label{fig:bias}
\end{figure}
\begin{figure}
    \includegraphics[width=\linewidth]{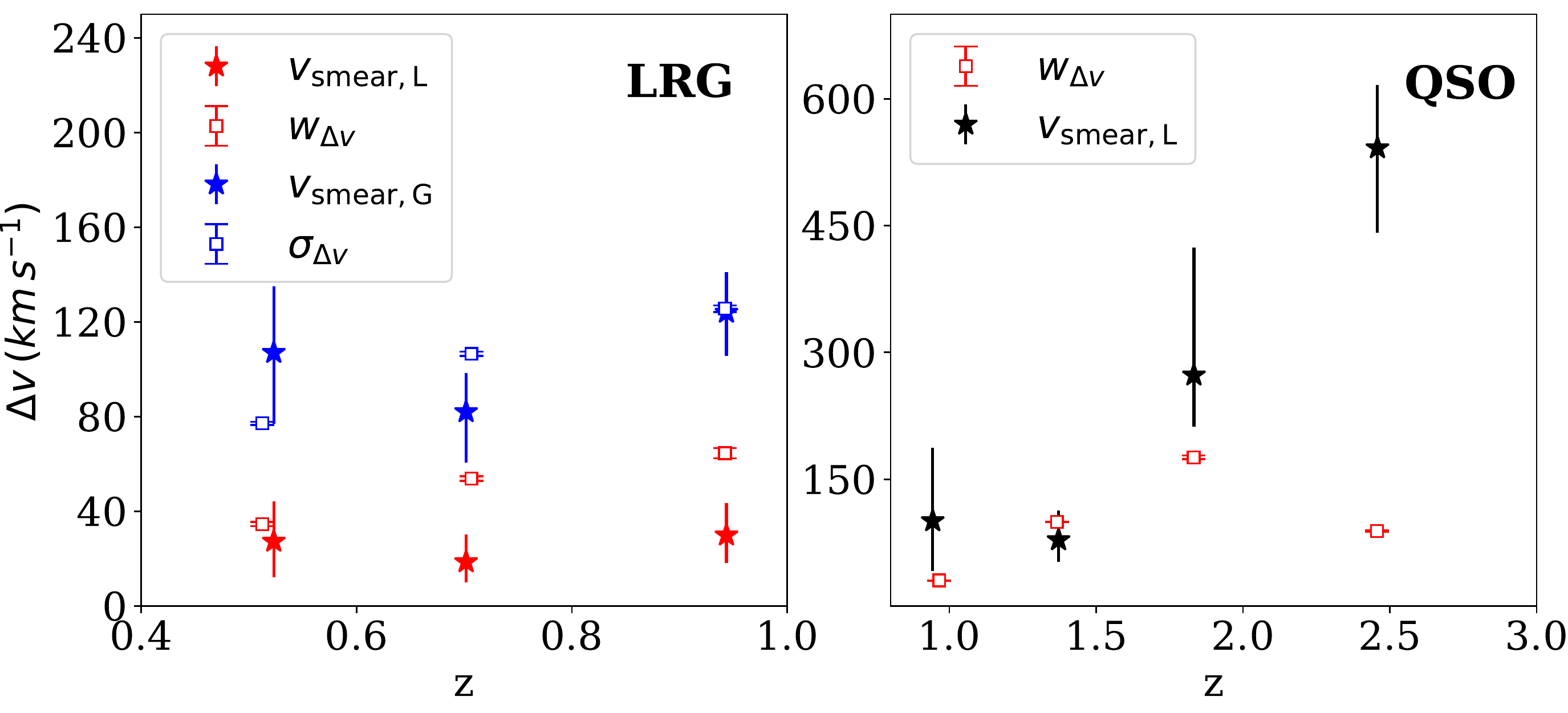}
    \caption{The redshift uncertainty quantified by the best-fitting SHAM $v_{\rm smear}$ (filled stars with error bars, the subscripts `L' and `G' stand for Lorentzian and Gaussian profiles) and that estimated statistically by the repeat observation $\Delta v$ (empty squares with error bars, $w_{\Delta v}$ for Lorentzian and $\sigma_{\Delta v}$ for Gaussian) for galaxies in different redshift slices. The results of Lorentzian profiles are in red colours and those of Gaussian profiles are in blue. For DESI LRGs, SHAM $v_{\rm smear,L}$ are systematically lower than $w_{\Delta v}$. In the case of the Gaussian profile, the SHAM $v_{\rm smear,G}$ and $\sigma_{\Delta v}$ (both with vertical offsets) agree with each other. For QSOs, the statistical uncertainty $w_{\Delta v}$ is not consistent with the SHAM $v_{\rm smear,L}$ (black filled stars with error bars) at $z>1.5$. 
    }
    \label{fig:deltav evolution}
\end{figure}
\begin{figure}
	\includegraphics[width=\columnwidth]{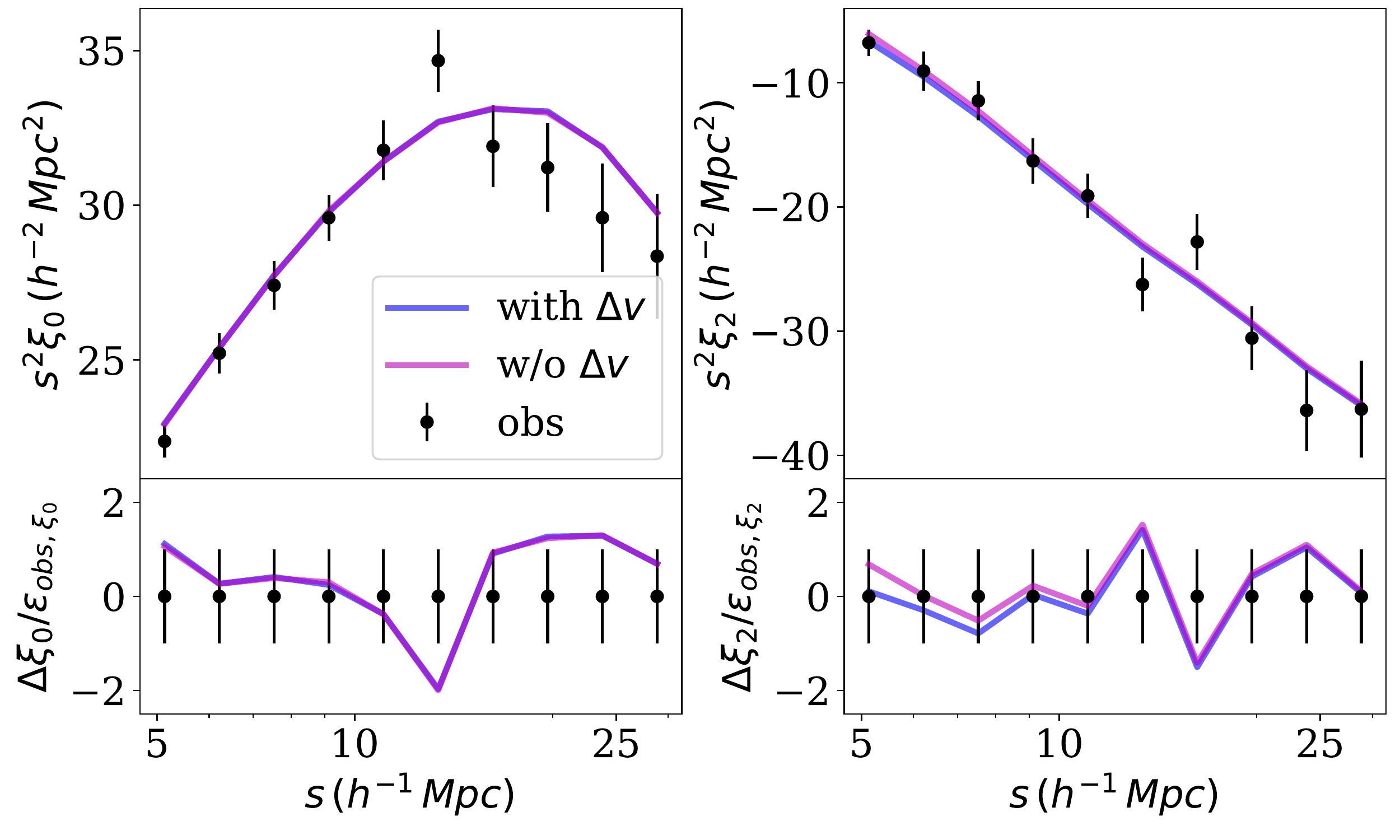}
    \caption{The effect of the maximum ELG redshift uncertainty $w_{\Delta v}=13.4\kms$ on the 2PCF monopole (left) and quadrupole (right). Multipoles of SHAM galaxies without $v_{\rm smear,L}$ are in magenta lines, the ones with $v_{\rm smear,L}=13.4\kms$ are in blue lines. Data are plotted in filled circles with error bars and the residuals normalized by the observed error bars are presented in the second row. No significant clustering effect is induced by the largest redshift uncertainty of ELGs at 5--30$\hmpc$. 
    }
    \label{fig:elg small effect}
\end{figure}
\begin{figure*}
    \includegraphics[width=\linewidth]{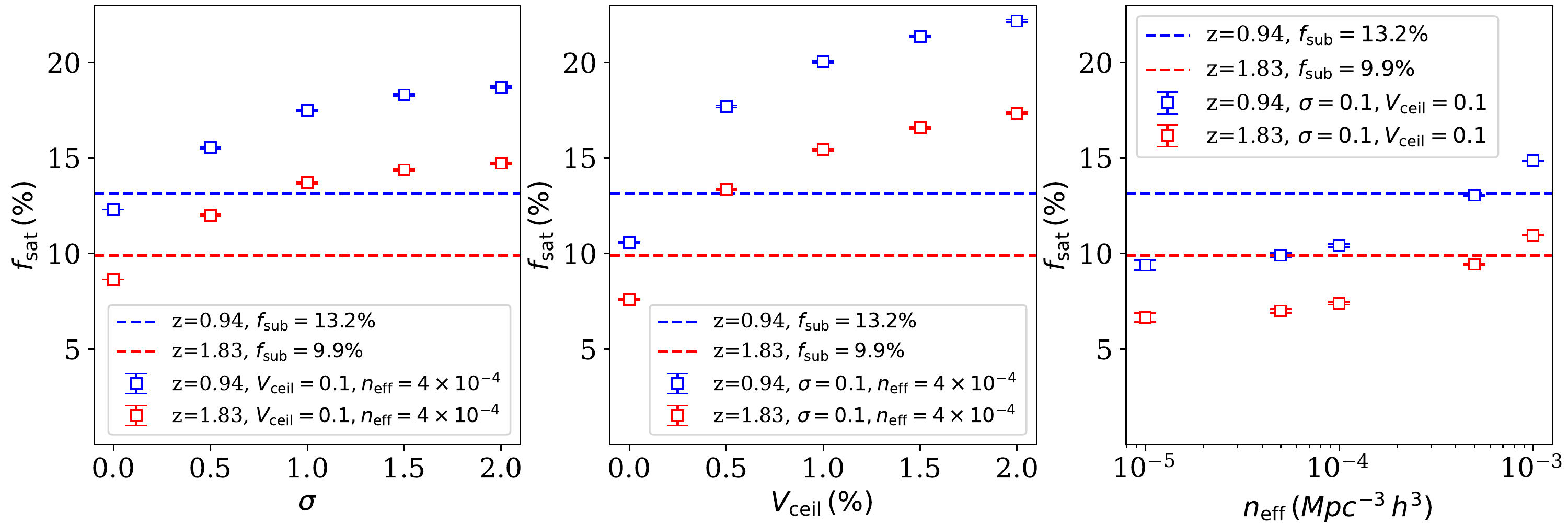}
    \caption{$f_{\rm sat}$ evolution with $\sigma$ (left panel), $V_{\rm ceil}$ (middle panel) and $n_{\rm eff}$ (right panel) for model galaxies of SHAM at $z=0.94$ (red empty error bars) and those at $z=1.83$ (blue empty error bars) produced by $v_{\rm smear}$-SHAM. We present the dependence of $f_{\rm sat}$ with one parameter and fix the other two parameters (typical values for LRGs) as indicated in the label. The error bar of $f_{\rm sat}$ is the standard deviation of $f_{\rm sat}$ among 32 realizations. The subhalo fraction of the UNIT simulation at $z=0.94$ (red) and $z=1.83$ (blue) defined in \refeq{eq:fsub} is plotted in dashed lines.
    }
    \label{fig:fsat impact zevol}
\end{figure*}
\subsection{Clustering}
\label{subsec:clustering}
We fit the 2PCF multipoles of the LRG, ELG and QSO samples from the DESI One Percent Survey at scales of 5--30$\hmpc$ over the redshift range $0.4<z<3.5$ with our SHAM algorithms. Table~\ref{tab:3param result} summarises the best-fitting parameters and their corresponding 1-$\sigma$ confidence intervals, as well as the minimum $\chi^2$ divided by the number of degrees of freedom. Note that $v_{\rm smear}$-SHAM results presented in Table~\ref{tab:3param result}, Figures~\ref{fig:lrg 2pcf}--\ref{fig:qso 2pcf} and in the appendices all use a Lorentzian $v_{\rm smear,L}$. In this case, $f_{\rm sat}$ is obtained as a derived parameter from the Monte-Carlo chain. 

Figures~\ref{fig:lrg 2pcf}--\ref{fig:qso 2pcf} are the 2PCF monopole (first row) and quadrupole (third row) of the observed tracers (filled circles with error bars) and model galaxies/QSOs generated using SHAM with the minimum-$\chi^2$ parameter set (solid lines). The shaded area around the SHAM clustering is the standard deviation of 2PCFs for all 32 SHAM realizations divided by $\sqrt{32}$. The observed error-rescaled residuals are presented in the second (monopole) and fourth (quadrupole) rows. 

The observed clustering of LRG samples is well-fitted by $v_{\rm smear}$-SHAM as shown in \reffig{fig:lrg 2pcf}. The reduced $\chi^2$ value of LRG fitting at $0.8<z<1.1$ is around $1.7$, which could be explained by under-fitting. However, our SHAM LRGs at $0.8<z<1.1$ also reproduce $w_p$ at $5<r_p<30\hmpc$ (see Appendix~\ref{appendix:reproduced wp} for the consistent projected 2PCF of SHAM LRGs and observations at this redshift bin). We attribute this large value to the off-diagonal terms in its jackknife covariance matrix. SHAM LRGs at $0.4<z<0.8$ have an underestimated $\xi_2$ at $r>20\hmpc$. At these scales, observations present a plateau while the quadrupoles of SHAM LRGs decrease. This flat quadrupole pattern is also present at $20-40\hmpc$ for LRGs from both SDSS-BOSS SGC at $z>0.4$ and eBOSS LRGs at all redshift bins, even after eliminating all known observational systematics \citep{2017systematics,Zhao2020}. The observed quadrupole resumes the smooth trend indicated by models at $s>50\hmpc$. Thus, the observed plateau could be attributed to cosmic variance or some uncorrected systematics for LRGs at $z>0.4$.
The SHAM underestimation may also indicate some shortcomings in our current understanding of the relationship between (sub)haloes and LRGs as was found in many BOSS and eBOSS galaxy mocks \citep[e.g.,][]{kitaura_clustering_2016,Rodriguez-Torres2016,Zhao2020,mine}, in particular for red galaxies \citep{BOSS_blue-red}. A detailed investigation of this problem is left for future work. 

ELG multipoles are well reproduced by SHAM galaxies at all redshift ranges as shown in \reffig{fig:elg 2pcf} and indicated by the reduced $\chi^2$ values in Table~\ref{tab:3param result}. 
QSO clustering has large observed errors due to the small number density of QSOs, which leads to large shot noise. \reffig{fig:qso 2pcf} proves that $v_{\rm smear}$-SHAM provides a consistently good description of the observation of QSOs in a large redshift range from $z=0.8$ to $z=3.5$. 

With the best-fitting catalogues of SHAM galaxies/quasars, we calculate their power spectrum with \textsc{pypowspec}\footnote{\url{https://github.com/dforero0896/pypowspec}} for the linear bias $b_{\rm lin}$ via
\begin{equation}
\label{eq:b_lin}
    b_{\rm lin}(z) 
    = \Bigl \langle \Bigl( \frac{P_{\rm g}(k,z)}{P_{\rm m}^{\rm lin}(k,z=0)}\Bigr) ^\frac{1}{2}\frac{1}{D(z)}\Bigr \rangle_{k<0.05\,h\,{\rm Mpc}^{-1}} 
\end{equation}
where $P_{\rm g}(k,z)$ is the power spectrum of SHAM galaxies at redshift $z$, $P_{\rm m}^{\rm lin}(k,z=0)$ is the linear matter power spectrum used by UNIT simulations renormalized to $z=0$. Both power spectra are in real space. $D(z)$ is the linear growth rate at redshift $z$. $k <0.05\,h\,{\rm Mpc}^{-1}$ in \refeq{eq:b_lin} means that the result is obtained by averaging over this $k$ range. 

\reffig{fig:bias} presents $b_{\rm lin}$ of LRGs (red), ELGs (blue) and QSOs (yellow) at different redshifts. The error bars are the weighted standard deviation of the linear bias for SHAM galaxies in the Monte-Carlo chain. $b_{\rm lin}$ increases with the redshift for each tracer. This is because we have a constant magnitude cut \citep{DESICollaboration2016a}, thus we can observe more low-luminosity galaxies/quasars at low redshift compared to the case at high redshift, resulting in an increasing bias with respect to the redshift. Studies using the same DESI EDR data show similar trends and consistent values for $b_{\rm lin}$ \citep{Yuan2023,Rocher2023,Prada2023}.

\subsection{Scatter \texorpdfstring{$\sigma$}{sigma} in galaxy--halo mass relation }
\label{subsec:gh scatter}
As discussed in \citetalias{mine}, $\sigma$ in our $v_{\rm smear}$-SHAM is composed of the intrinsic scatter in the galaxy--halo mass relation and the completeness for galaxies with an intermediate stellar mass. For LRG samples, $\sigma\sim0.2$ dex. However, since there is a degeneracy between $\sigma$ and $V_{\rm ceil}$ (see Appendix~\ref{appendix:Posterior Contours} for the posteriors of LRGs), it is not clear whether there is a redshift evolution in $\sigma$. For ELG samples, the constraints in $\sigma$ are weak regardless of the prior range. Given its large number density, this is not the result of large errors in clustering, as is the case for QSOs. The (sub)halo incompleteness of ELGs is related to their incompleteness in stellar mass and in luminosity \citep{favole_clustering_2016,violeta_number_density}. This leads to a complex galaxy property--halo mass connection for ELGs, thus a weakly constrained $\sigma$ as it integrates many factors. 
$\sigma$ also leads to the stochastic variance in the clustering of SHAM galaxies. For the LRG and ELG samples, this variance is as small as 5 per cent of the observational error $\epsilon_{\rm obs}$ (not visible as shown in \reffig{fig:lrg 2pcf}--\ref{fig:elg 2pcf}). So we can also ignore its effect on the final $\chi^2$ values in general. 

\subsection{Massive (Sub)Halo Incompleteness \texorpdfstring{$V_{\rm ceil}$}{Vceil}}
\label{subsec:completeness}
For LRGs, $V_{\rm ceil}$ describes the halo/stellar mass incompleteness at the massive end. As shown in LRG posteriors (See Appendix~\ref{appendix:Posterior Contours}), $V_{\rm ceil}$ values for LRGs are very small but are definitely non-zero at 1-$\sigma$ level. This incompleteness can be attributed to the fact that the target selection of LRGs removes some of the most massive blue galaxies, leading to empty massive haloes. 

\begin{figure}
    \centering
	\includegraphics[width=\linewidth]{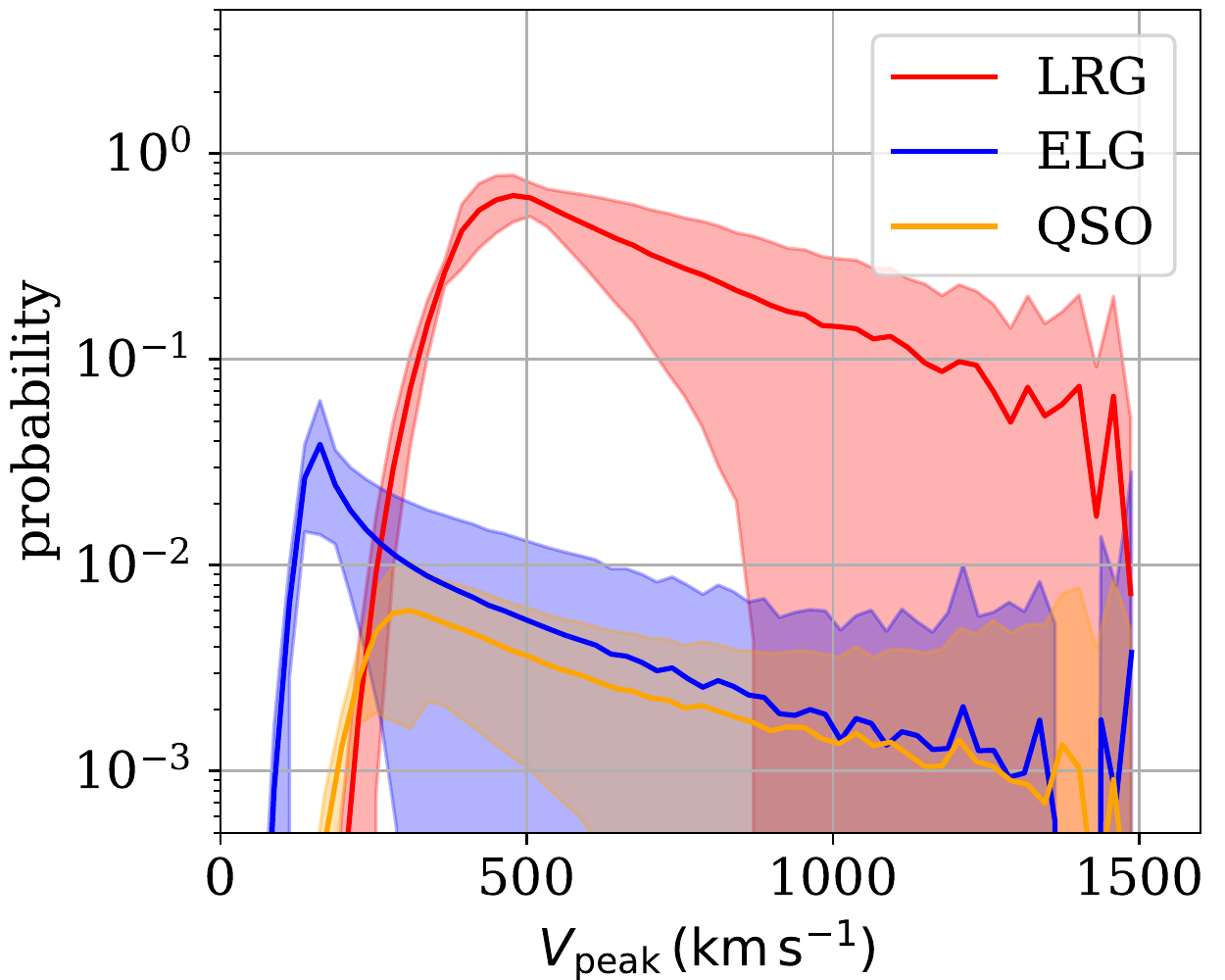}
    \caption{The probability of a (sub)halo to host an LRG (red), an ELG (blue) or a QSO (orange) as a function of $V_{\rm peak}$ of (sub)haloes for LRGs with Gaussian $v_{\rm smear,G}$, ELGs and QSOs at $0.8<z<1.1$. The shades are their 1-$\sigma$ errors calculated using the Monte-Carlo chain.}
    \label{fig:pdf}
\end{figure}

For ELG samples, $V_{\rm ceil}$ is critical to describe their absence in the centre of massive (sub)haloes. ELG entries in Table~\ref{tab:3param result} show that we shall remove a per-cent level of (sub)haloes, allowing few (sub)haloes with $V_{\rm peak}>300\kms$ to host an ELG at their centre (\reffig{fig:pdf}). 
Since there is no degeneracy among parameters of the $f_{\rm sat}$-SHAM, the $1-\sigma$ difference between the $V_{\rm ceil}$ of ELGs at $0.8<z<1.1$ and $1.1<z<1.6$ embodies their clustering difference.  

QSOs at $0.8<z<1.1$ show a larger $V_{\rm ceil}$ than that of QSOs at higher redshifts, but this difference is not significant due to the weak constraint. This is consistent with \citet{DESI_QSO_TS} in which QSOs at $z<1.0$ show a smaller purity than those at high redshifts. Even though $V_{\rm ceil}$ for QSOs only exclude less than 1 per cent of (sub)haloes with the largest $V_{\rm scat}$ (\refeq{eq:Vpeak_scat}), the clustering of QSOs cannot be well-fitted without the $V_{\rm ceil}$ parameter. 
As explained in Section~\ref{subsec:sham}, this is consistent with findings of SAM studies \citep[e.g.,][]{griffin2019} and observations \citep[e.g.,][]{QSOincomplete2017}, in which QSOs are absent in the centre of the most massive (sub)haloes. 

\begin{figure}
    \centering
	\includegraphics[width=\linewidth]{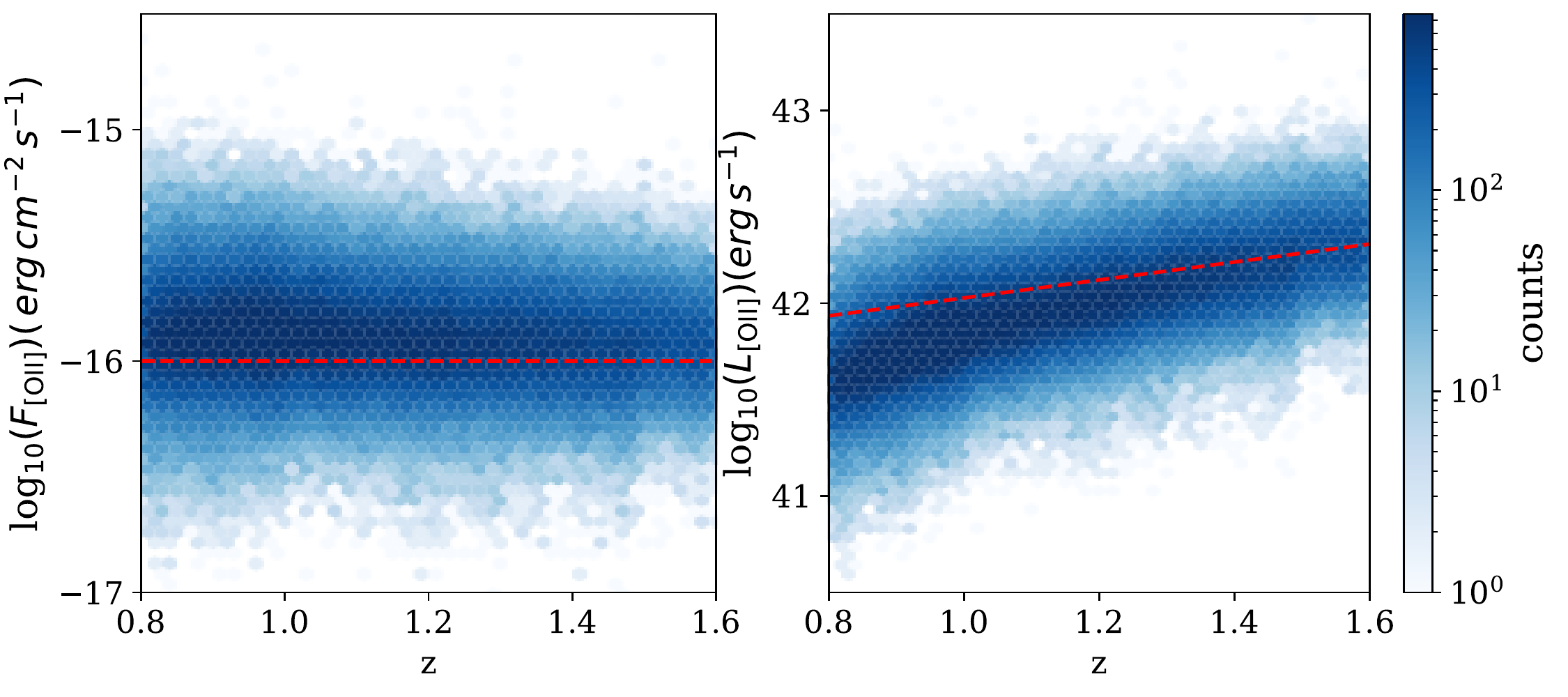}
    \caption{The flux (in $\rm erg\,cm^{-2}\,s^{-1}$) and the luminosity (in $\rm erg\,s^{-1}$) of \oii{} emission for ELGs as a function of the redshift. The red lines are the strong \oii{} emitter threshold. The line on the left represents the standard of \citet{lowfsat} with $F_{\rm \oii}>10^{-16}\ergs$. The one on the right shows an evolving $L_{\rm \oii}^{\rm thres}$ (\refeq{eq:oii thres}) derived from \citet{Gao2022}. 71 per cent of DESI ELGs pass the $F_{\rm \oii}$ selection, 24 per cent pass the $L_{\rm \oii}$ selection.}
    \label{fig:oii threshold}
\end{figure}
\begin{figure}
    \centering
	\includegraphics[width=\linewidth]{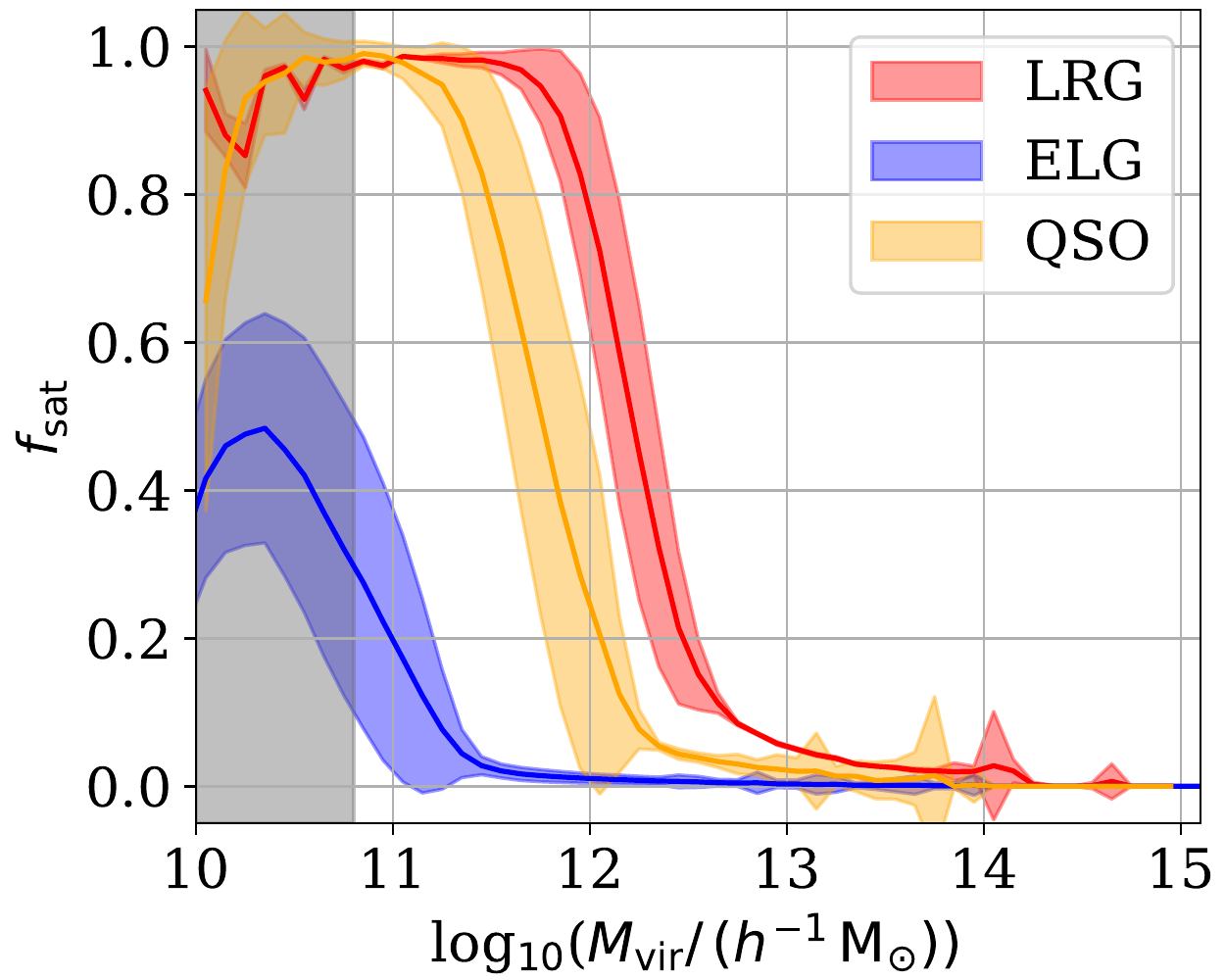}
    \caption{The $f_{\rm sat}$-$M_{\rm vir}$ relation of SHAM LRGs at $0.4<z<1.1$ (red), ELGs at $0.8<z<1.6$ (blue) and QSOs at $0.8<z<3.5$ (orange). The shaded areas around solid lines indicates the 1-$\sigma$ errors of the $f_{\rm sat}$-$M_{\rm vir}$ relation are derived from the Monte-Carlo chain. The grey areas show $M_{\rm vir}<6\times10^{10}\Msun{}$ (Section~\ref{subsec:sim}). }
    \label{fig:fsat_Mvir}
\end{figure}
\begin{figure*}
    \centering
	\includegraphics[width=\linewidth]{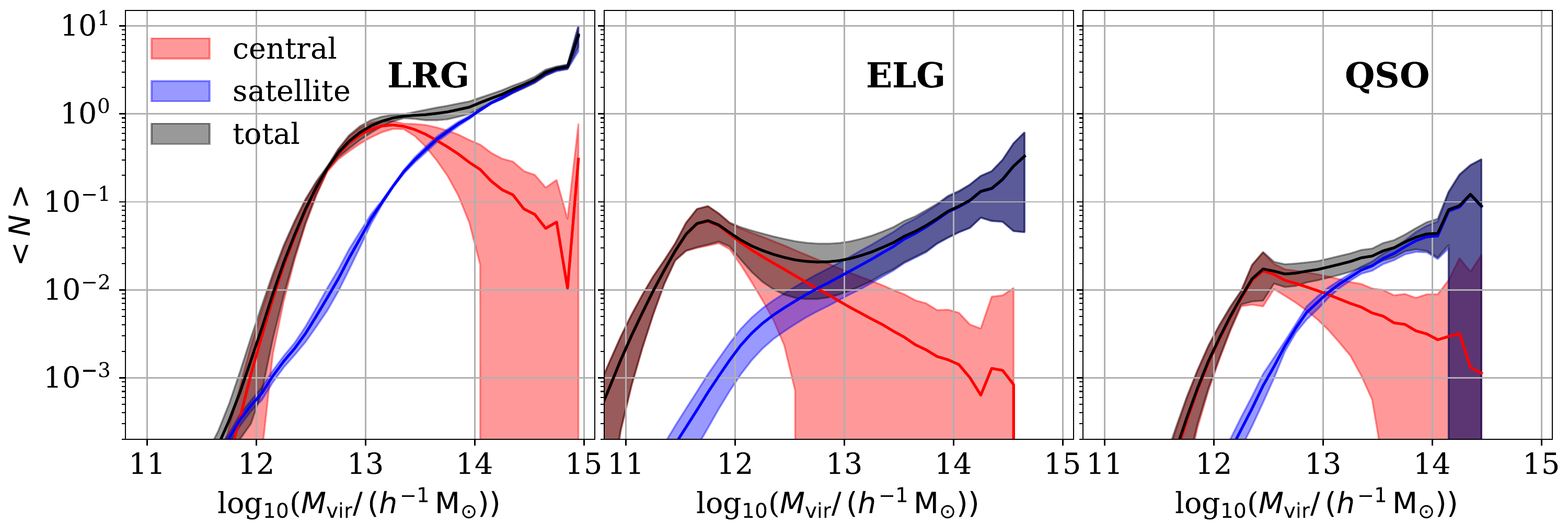}
    \caption{The average HOD of SHAM LRGs at $0.4<z<1.1$ (left panel), ELGs at $0.8<z<1.6$ (middle panel) and QSOs at $0.8<z<3.5$ (right panel). The contribution of central galaxies to the average HOD is shown by red lines and that of satellites by blue lines. The 1-$\sigma$ errors derived from the Monte-Carlo chain are shown as shaded regions of the same colour. }
    \label{fig:hod}
\end{figure*}
\begin{figure}
    \centering
	\includegraphics[scale=0.5]{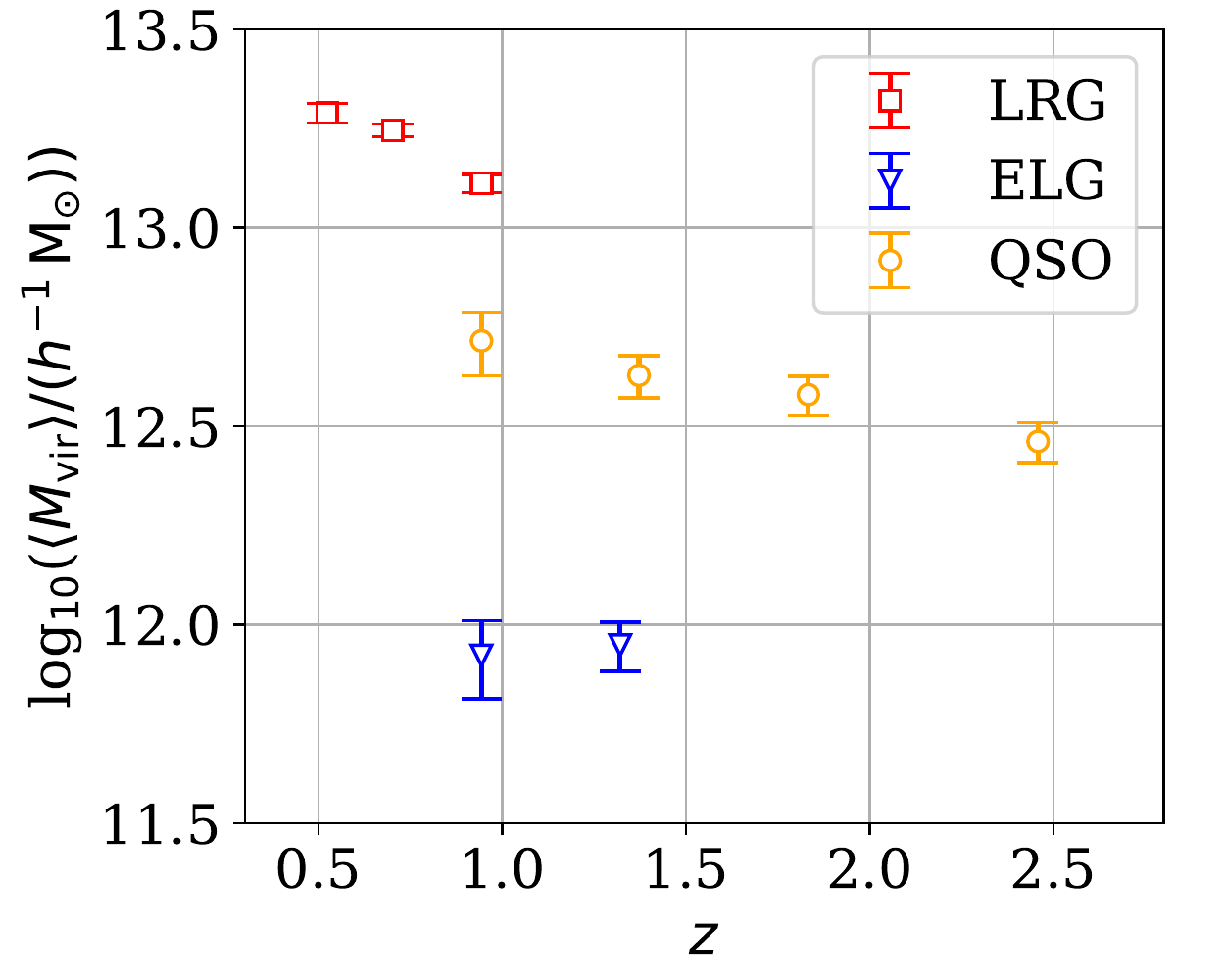}
    \caption{The evolution of the mean parent halo mass $\langle M_{\rm vir}\rangle$ of SHAM LRGs with Gaussian $v_{\rm smear,G}$ (red empty square with error bars), ELGs (blue empty triangle with error bars) and QSOs (orange empty circle with error bars) obtained for sub-samples at redshift slices as a function of redshift. There are three ranges of mean mass and the values of LRGs and QSOs decrease with redshift. 
    }
    \label{fig:meanMvir}
\end{figure}
\subsection{Redshift Uncertainty \texorpdfstring{$v_{\rm smear}$}{Vsmear}}
\label{subsec:vsmear and deltav}
$v_{\rm smear}$-SHAM uses a Lorentzian $v_{\rm smear,L}$ by default as this is a good model for DESI tracers in general. In particular, for LRGs, Lorentzian and Gaussian profiles can both describe the redshift difference distribution of LRG repeat observations (\reffig{fig:deltav measurement}). So we perform $v_{\rm smear}$-SHAM fitting with Gaussian $v_{\rm smear,G}$ as well and Table~\ref{tab:gaussian result} includes the best-fitting results. The $\chi^2$ values of SHAM with $v_{\rm smear,G}$ are similar to those of SHAM with $v_{\rm smear,L}$. Their best-fitting $\sigma$ and $V_{\rm ceil}$ are also consistent with each other, meaning that the clustering effect of a truncated Lorentzian $v_{\rm smear,L}$ and a Gaussian $v_{\rm smear,G}$ is almost equivalent. So we need to check the $v_{\rm smear}$ consistency with the statistical redshift uncertainty estimated from repeat observations. 

\reffig{fig:deltav evolution} is the comparison between the best-fitting SHAM $v_{\rm smear}$ (filled stars with error bars) and the best-fitting redshift uncertainty measured from repeat observations (empty squares with error bars) for DESI LRGs (left panel) and QSOs (right panel). The results of Lorentzian profiles are in red and those of Gaussian profiles are in blue. In particular, Gaussian profiles for LRGs are vertically shifted. The best-fitting LRG SHAM $v_{\rm smear,G}$ values agree with the Gaussian dispersion $\sigma_{\Delta v}$ of repeat observations, while SHAM $v_{\rm smear,L}$ values underestimate the statistical redshift uncertainty evaluated using the width of Lorentzian functions $w_{\Delta v}$. Note that the standard deviation $\hat{\sigma}_{\Delta v}$ of LRG repeats are also consistent with the dispersion of the Gaussian profile. Therefore, the Gaussian profile $v_{\rm smear,G}$ is more suitable for illustrating the uncertainty of the redshift of LRGs in redshift bins. 

In contrast, the Lorentzian profile results ($w_{\Delta v}=46.5\pm0.8\kms$ while $v_{\rm smear,L}=40\pm9\kms$) are in better agreement for LRGs in the full redshift range, indicating that there are multiple types of LRGs with different redshift uncertainty properties. As a result, \citet{DESI_ELG_TS, DESI_QSO_TS} use a linear combination of Gaussian profiles to fit the redshift difference $\Delta v$. Nevertheless, a truncated Lorentzian with just one parameter works in the same way as multiple-Gaussian profiles in terms of the clustering effects and SHAM results.

For QSOs in the right panel of \reffig{fig:deltav evolution}, SHAM $v_{\rm smear,L}$ shows a non-decreasing trend. This reflects the quadrupole amplitude of QSOs at different redshift bins shown in the third row of \reffig{fig:qso 2pcf}. However, this trend is inconsistent with that of repeat observations (red empty error bars), and SHAM $v_{\rm smear,L}$ starts to deviate from $w_{\Delta v}$ of repeat observations at $z\gtrsim 1.5$. The discrepancy is possibly due to the switch of the main spectral line for redshift determination from Mg\,\textsc{ii} and C\,\textsc{iv} \citep{Zarrouk2018}. QSO spectral lines are subject to systematical velocity shifts caused by astrophysical effects \citep[e.g.,][]{qsoshift1982,qsoshift2002,qsoshift2011} and repeat observations for the same object cannot capture this shift. But this shift is different from object to object, creating an extra relative random motion between QSO pairs. Moreover, \citet{Shen2016} have proved that Mg\,\textsc{ii} is the least shifted broad emission of QSOs, while C\,\textsc{iv} can be strongly shifted. This will result in a larger random motion between QSO pairs at $z>1.5$ than those at $z<1.5$. This motion is integrated into $v_{\rm smear, L}$ of SHAM, resulting in its fast rise at $z>1.5$ and explaining the inconsistency between $v_{\rm smear, L}$ and $w_{\Delta v}$.

The redshift uncertainty of ELGs is small in general as presented in \reffig{fig:deltav measurement}. Its largest redshift uncertainty among ELGs samples, i.e., $v_{\rm smear,L}=13.4\kms$ for ELGs at $1.1<z<1.6$, does not produce a significant clustering effect, as illustrated in \reffig{fig:elg small effect}. The magenta line is the clustering of SHAM galaxies created using the best-fitting $f_{\rm sat}$-SHAM. The blue line shows the clustering with $v_{\rm smear,L}=13.4\kms$ applied to the peculiar velocity of the SHAM ELGs. For the monopole, $v_{\rm smear,L}$ have little influence as expected, and the influence on the quadrupole is restricted to 5--8$\hmpc$ and is within 1-$\sigma$ range of the observed jackknife error bars. Therefore, asserting $v_{\rm smear}=0$ for $f_{\rm sat}$-SHAM does not bias the best-fitting results of the other three parameters.

\subsection{Satellite Fraction \texorpdfstring{$f_{\rm sat}$}{fsat}}
\label{subsec:satellite fraction}
The fraction of satellites, $f_{\rm sat}$, for LRGs and QSOs, is calculated when $\sigma$ and $V_{\rm ceil}$ are given to $v_{\rm smear}$-SHAM. We present in \reffig{fig:fsat impact zevol} the impact of $\sigma$, $V_{\rm ceil}$ and the number density $n_{\rm eff}$ on $f_{\rm sat}$ for SHAM galaxies produced by $v_{\rm smear}$-SHAM at $z=0.94$ (blue) and $z=1.83$ (red). We $f_{\rm sub}$ of UNIT simulations at those redshifts are plotted in dashed horizontal lines, obtained using \refeq{eq:fsub}. The fixed parameters\footnote{These are typical values for LRGs. We have checked the $f_{\rm sat}$ relations with typical values for QSOs and found the same trends as in \reffig{fig:fsat impact zevol}.} are $\sigma=0.1$, $V_{\rm ceil}=0.1$, $n_{\rm eff}=4\times10^{-4}\,{\rm Mpc}^{-3}\,h^3$. $f_{\rm sat}$ monotonically increases with $\sigma$, $V_{\rm ceil}$ and $n_{\rm eff}$. This is because larger $\sigma$, $V_{\rm ceil}$ and $n_{\rm eff}$ all mean selecting more (sub)haloes with small $V_{\rm peak}$, corresponding to a larger fraction of subhaloes/satellites. In contrast, for $\sigma,V_{\rm ceil}\sim 0$, there are a few selected subhaloes that have a large $V_{\rm peak}$, resulting in $f_{\rm sat}<f_{\rm sub}$. Due to the tight constraints on LRG $\sigma$ and $V_{\rm ceil}$, the 1-$\sigma$ confidence interval of LRG $f_{\rm sat}$, which is a derived parameter, is also small. From another aspect, the slope of the $f_{\rm sat}$--$\sigma$ and $f_{\rm sat}$--$V_{\rm ceil}$ relations decreases as $\sigma$ and $V_{\rm ceil}$ become larger. Additionally, the slope of the $f_{\rm sat}$--$n_{\rm eff}$ relation increases with $n_{\rm eff}$. The combination of these effects results in the small errors of QSO $f_{\rm sat}$ despite its loose constraints on $\sigma$ and $V_{\rm ceil}$. 

The satellite fraction of $v_{\rm smear}$-SHAM is also affected by the redshift, i.e., the substructure growth. $f_{\rm sat}$ of SHAM galaxies at $z=1.83$ are lower than those at $z=0.94$, calculated with the same $v_{\rm smear}$-SHAM parameters. This is consistent with the decreasing trend of $f_{\rm sat}$ for LRGs and QSOs with the redshift.

For ELGs, we find that about $4$ per cent of them are satellite galaxies when we fit the data with our $f_{\rm sat}$-SHAM method. Such a low fraction of satellites is also found in models mimicking a DESI-like survey~\citep{lowfsat}. In the literature, for different selections of ELGs, this fraction has been found to range from $f_{\rm sat}\sim 5$ to $\sim 22.5$ per cent~\citep{favole_clustering_2016,Gao2022} in VIPERS and from $\sim 17$ per cent~\citep{eboss_elg_fsat} to 19.3 per cent~\citep{eboss_elg_sham} in eBOSS. The difference in the strength of the \oii{} emission can also alter $f_{\rm sat}$. For example, \citet{lowfsat,Gao2022} find that strong \oii{} emitters tend to have a low $f_{\rm sat}$ down to per cent level, i.e., 4.6 and $7.0\pm2.0$ per cent respectively.
In \reffig{fig:oii threshold} we show the distribution of DESI ELG \oii{} fluxes, $F_{
\rm \oii}$, and luminosities, $L_{\rm \oii}$, as a function of redshift. We find that more than $71$ per cent of DESI ELGs from the One Percent Survey have $F_{\rm \oii}>10^{-16}\ergs$ (the red line on the left panel), which is the cut assumed in the theoretical study of \citet{lowfsat}. On the right panel, $24$ per cent of them are strong \oii{} emitters according to the definition in \citet{Gao2022}, i.e., they should have $L_{\rm \oii}$ larger than
\begin{equation}
    {\rm log}_{10} L_{\rm \oii}^{\rm thres}(z) = {\rm log}_{10} L_{\rm \oii}^{\rm thres}(z=0.5)+{\rm log}_{10} \left( \frac{1+z}{1+0.5} \right) ^{\beta_{L}}
    \label{eq:oii thres}
\end{equation}
where $z\in (0.8,1.6)$, and $L_{\rm \oii}^{\rm thres}(z=0.5)=10^{41.75}\,\rm erg\,s^{-1}$, which is the $L_{\rm \oii}$ lower bound of VIPERS ELGs with $f_{\rm sat}=7.0\pm 2.0$ per cent. This fraction for VIPERS and eBOSS are 10 and 12 per cent respectively. Thus, it is reasonable if we obtain a smaller $f_{\rm sat}$ from the DESI data, compared to that of the eBOSS ELGs and VIPERS samples.

However, it is still possible that our satellite fraction is underestimated as we do not include orphan galaxies that are necessary for correcting the deficits of the current subhalo tracking method \citep{Behroozi2019}. We check in Appendix~\ref{appendix:fsat bias} the consistency between the $f_{\rm sat}$ measured by $f_{\rm sat}$-SHAM and galaxy mocks provided by UniverseMachine \citep{Behroozi2019} and SAM models \citep{lowfsat}. The mass resolution of the UNIT simulation may also not be good enough to resolve all substructures for ELGs. Moreover, $f_{\rm sat}$ is model dependent~\citep[e.g.,][]{favole_clustering_2016,Gao2022}. In studies for DESI ELGs, \citet{Gao2023} present a redshift- and stellar-mass-dependent satellite fraction with reconstructed orphan galaxies. \citet{Rocher2023} find that adding conformity can lead to a smaller $f_{\rm sat}$ compared to HOD without that. So it is difficult to compare fairly the $f_{\rm sat}$ value provided by different models for different galaxy surveys. We will leave those for future work.

In \reffig{fig:fsat_Mvir}, we present the satellite fraction $f_{\rm sat}$ as a function of the (parent) halo mass $M_{\rm vir}$. For LRGs and QSOs, almost all galaxies residing on small haloes selected by SHAM are satellites, and then $f_{\rm sat}$ decreases to 0 as $M_{\rm vir}$ increases to $8.9\times 10^{13}\Msun{}$ for LRGs and $1.8\times10^{13}\Msun{}$ for QSOs. For ELGs, less than half of the selected small haloes host satellites, then the satellite fraction decreases to 0 as $M_{\rm vir}>2.2\times10^{12}\Msun{}$.

\subsection{Halo Occupation Distribution}
\label{subsec:hod}
\reffig{fig:hod} shows the average halo occupation distribution (HOD) of SHAM LRGs with Lorentzian $v_{\rm smear,L}$ at $0.4<z<1.1$, SHAM ELGs at $0.8<z<1.6$ and SHAM QSOs at $0.8<z<3.5$ as a function of the halo mass $M_{\rm vir}$. They are computed with the weights from the Monte-Carlo chain.
$M_{\rm vir}$ corresponds to the virial mass of host haloes or parent haloes of subhaloes. We opt not to compare directly our halo occupation with the other DESI EDR galaxy-halo connection results. This is because our HODs originate from different N-body simulations, varying in redshift, halo finders, and mass resolution. These differences might influence the HOD configuration. Consequently, rather than pursuing a direct comparison of HOD, our target in the following section is to conclude the common features of DESI EDR tracers measured by different galaxy-halo connection methods and the characteristics of different tracers provided by our SHAM. 

For LRGs, their stellar mass is closely related to the halo mass \citep[e.g.,][]{leauthaud2012,Behroozi2019}. Their HOD can be modelled using a 5-parameter form, with a smoothed step function for central galaxies, and a power law for satellites \citep[e.g.,][]{HOD2005,eboss_lrg_hod2017}.
The HOD of our central SHAM LRGs reaches $\langle N\rangle=0.75\pm0.07$ at $M_{\rm vir}=10^{13.3}\Msun{}$ and decreases towards the massive end as we set $V_{\rm ceil}$ free. This incompleteness is consistent with the measurement of \absum HOD for DESI LRGs \citep{Yuan2023}. 

There are various HOD models for central ELG. \citealp{eboss_elg_hod}; \citet{Rocher2023} discuss several ELG central profiles: the modified high-mass-quenched model \citep{HMQm2020}, the Gaussian function, the star-forming HOD model \citep{eboss_elg_hod}, the lognormal HOD model \citep{Rocher2023}. The central HOD of our SHAM ELGs shows a preference for a star-forming HOD profile with a turning point at $M_{\rm vir}=10^{11.7}\Msun{}$ that reaches $\langle N\rangle=0.06\pm0.03$. ELGs residing in $M_{\rm vir}>10^{12.5}\Msun{}$ are also found in the study of \citet{Gao2023}. 

There are also multiple profiles for QSO HOD models \citep{eboss_qso_hod_2020,Yuan2023}. Our central QSO HOD reaches the maximum value $\langle N\rangle=0.016\pm0.001$ after $M_{\rm vir}=10^{12.4}\Msun{}$. Note that no tracer reaches $\langle N\rangle=1$. It means that we will not find one galaxy/QSO in every halo above a certain halo mass. It is consistent with the $V_{\rm ceil}$ results that LRGs from the One-Percent Survey are not complete, while ELGs and QSOs are absent from massive haloes due to physical reasons.

Note that our SHAM model galaxies/quasars present a decreasing number of centrals in the massive halo in \reffig{fig:hod} (red lines with shades). This is because we apply a simple, empirical truncation to the massive haloes via $V_{\rm ceil}$, aiming at recover the auto-correlations of the DESI EDR tracers with a minimum number of parameters. Therefore, the increasing incompleteness at the massive end for model galaxies is not necessarily physical, given the lack of assembly bias effect for example. In fact, \citet{Rocher2023} find a similar decreasing trend for central ELGs in DESI with four different models that include the assembly bias. Meanwhile, \citet{Yuan2023} provide a constant number of central galaxies/quasars in massive haloes and find incompleteness there for both LRGs and QSOs from DESI, with the HOD models including the assembly bias. Nevertheless, the fact that all these galaxy/quasar-halo connection models show a central galaxy occupation below unity regardless of the assembly bias suggests that LRGs, ELGs, and QSOs from the DESI One-Percent Survey are likely incomplete in their host halo masses. We will need a more sophisticated galaxy-halo relation as well as better observations and simulations to understand this incompleteness better. 

The average HOD of our SHAM satellites for all tracers can be fitted by two exponential functions of $M_{\rm vir}$, i.e., 
\begin{equation}
\label{eq:hod sat}
\langle N_{\rm sat} \rangle \propto
    \begin{cases}
        M_{\rm vir}^\alpha, & M_{\rm vir} > M_{\rm turn}.\\
        \\
        M_{\rm vir}^\beta, & M_{\rm vir} < M_{\rm turn}.
    \end{cases}
\end{equation}
The best-fitting results obtained by \textsc{PyMultinest} are presented in Table~\ref{tab:HODparam}. The second exponent $\beta$ at a lower mass range for all tracers is consistent and around 2. Though $M_{\rm turn}$ values are different for different tracers, their slope in the massive end $\alpha$ is well consistent with 0.7. This is consistent with ELG HOD slopes in \citet{Rocher2023} but smaller than those from LRG HOD in \citet{Yuan2023}.
\begin{table}
    \scalebox{0.85}{
    \begin{tabular}[c]{c|c|c|c|c|c}
    \hline 
    \hline 
    {tracer}&{$\alpha$}&{$\beta$}&{$\text{log}_{10}(M_{\rm turn})$}&{$\text{log}_{10}(\langle M_{\rm vir}\rangle)$}&{$\langle V_{\rm peak}\rangle$}\\
    {}&{}&{}&{$(\Msun{})$}&{$(\Msun{})$}&{$(\kms)$}\\
    \hline
    {LRG}&{${0.70}^{+0.01}_{-0.01}$}&{${1.97}^{+0.01}_{-0.01}$}&{${13.55}^{+0.00}_{-0.00}$}&{${13.18\pm0.01}$}&{$457\pm6$}\\
    \hline
    {ELG}&{${0.76}^{+0.06}_{-0.07}$}&{${2.16}^{+0.13}_{-0.14}$}&{${12.05}^{+0.20}_{-0.20}$}&{$11.90\pm0.06$}&{$159\pm4$}\\
    \hline
    {QSO}&{${0.73}^{+0.07}_{-0.07}$}&{${2.13}^{+0.05}_{-0.05}$}&{${12.95}^{+0.10}_{-0.10}$}&{${12.59\pm0.03}$}&{$346\pm8$}\\
    \hline    
    \end{tabular}
    }
    \caption{The best-fitting $\alpha$, $\beta$, $M_{\rm trun}$ of $\langle N_{\rm sat} \rangle$, the mean haloe mass $\langle M_{\rm vir}\rangle$ and the mean $V_{\rm peak}$ $\langle V_{\rm peak}\rangle$ for LRGs at $0.4<z<1.1$, ELGs at $0.8<z<1.6$ and QSOs at $0.8<z<3.5$.}
    \label{tab:HODparam}
\end{table}

The mean parent halo mass $\langle M_{\rm vir}\rangle$, derived from the Monte-Carlo chain as introduced in Section~\ref{subsec:fitting}, is shown in Table~\ref{tab:HODparam}. Those values are consistent with those of \absum HOD using the same data from the DESI One Percent Survey \citep{Yuan2023,Rocher2023}.
\reffig{fig:meanMvir} is the evolution of the mean parent halo mass for SHAM LRGs with Gaussian $v_{\rm smear,G}$, ELGs, and QSOs. $\langle M_{\rm vir}^{\rm LRG}\rangle$ is not smaller than $10^{13}\Msun{}$, and $\langle M_{\rm vir}^{\rm QSO}\rangle$ values range from $10^{12.2}\Msun{}$ to $10^{12.7}\Msun{}$. Both decrease with redshift, consistent with the redshift-evolution HOD results from \citet{Yuan2023}. For ELG parent halo, there is no significant evolution and both $\langle M_{\rm vir}\rangle$ are below $10^{12}\Msun{}$. \citet{Rocher2023} and \citet{Gao2023} also find the same feature for the mean halo mass of ELGs but with slightly different values. Those are all consistent with the expectation of \citet{DESICollaboration2016a}. But the bias of ELGs still increases with redshift as indicated in \reffig{fig:bias}. 
It should be noted that only 1.6 per cent of ELGs and less than 0.03 per cent of LRGs and QSOs reside in haloes lower than the reliable mass threshold $6\times10^{10}\Msun{}$ (Section~\ref{subsec:sim}). So the influence of the limitation in the $N$-body simulation halo finder on our SHAM study can be dismissed.

The PDF of (sub)halo $V_{\rm peak}$ that can host a central (satellite) galaxy is shown in \reffig{fig:pdf} for all three tracers at $0.8<z<1.1$. This is the empirical galaxy--halo relation that we calibrate using SHAM described in Section~\ref{subsec:sham}, i.e., LRG with Gaussian $v_{\rm smear,G}$, ELG and QSO. The shape of the PDF is modulated indirectly by $\sigma$ and $V_{\rm ceil}$. ELGs mainly reside in (sub)haloes with $V_{\rm peak}\lesssim200\,\kms$, while LRG and QSOs are populated in haloes with $V_{\rm peak}>200\,\kms$. The PDFs of LRGs and ELGs present a clear peak, while the probability of QSOs in the massive end decays slower than them. The probability patterns for different tracers can be used as a reference for future multi-tracer studies. The mean $V_{\rm peak}$ values for the total LRG, ELG and QSO samples are presented in Table~\ref{tab:HODparam}.



\section{Conclusions}
\label{sec:conclusion}
We have generated catalogues of mock galaxies matching the clustering of dark tracers from the DESI One Percent Survey in the range of 5--30$\hmpc$. The DESI samples studied here are luminous red galaxies (LRG) at $0.4<z<1.1$; emission line galaxies (ELG) at $0.8<z<1.6$ and QSO at $0.8<z<3.5$ (Section~\ref{subsec:EDR}). Mock galaxies are painted on the dark matter only UNIT simulation (Section~\ref{subsec:sim}), using two SubHalo Abundance Matching (SHAM) algorithms (Section~\ref{subsec:sham}). The first algorithm, $v_{\rm smear}$-SHAM, is used for LRGs and QSOs, and has the following free parameters: $\sigma$, to model the dispersion in the galaxy--halo mass relation but also include the incompleteness of halo mass; $V_{\rm ceil}$, to account for the incompleteness of massive haloes for the galaxy samples; and $v_{\rm smear}$, to model the uncertainty in the redshift determination process. The other SHAM model, $f_{\rm sat}$-SHAM with a free satellite fraction $f_{\rm sat}$, is introduced here to model ELGs as $f_{\rm sat}$ is crucial to recover the quadrupole of DESI ELGs. The redshift uncertainty of ELGs is the lowest among the considered tracers and its $v_{\rm smear}$ has a negligible impact on the clustering of SHAM ELGs down to 5$\hmpc$. So $v_{\rm smear}$ is not included in $f_{\rm sat}$-SHAM.

For LRGs, we find the best-fitting $\sigma$ to be consistent with 0 at a 2-$\sigma$ level. However, ELG and QSO samples constrain $\sigma$ weakly. Although the loose constraint from the QSO sample is mostly due to its small number density, this does not stand for ELGs which are over 10 times denser than QSOs. We attribute this lack of constraint to the fact that $\sigma$ also models the incompleteness in both stellar mass and luminosity, resulting in a complex galaxy--halo relation that is harder to constrain.

$V_{\rm ceil}$, the massive-(sub)halo incompleteness, describes the stellar mass incompleteness in the massive end due to both target selection criteria of galaxy surveys and the intrinsic properties of certain galaxies. The best-fitting $V_{\rm ceil}$ for LRGs, is as small as 0.02 per cent. This small value shows that DESI LRGs from the One Percent Survey are close to complete at the massive end. 
The best fit $V_{\rm ceil}$ for ELGs shows that up to 7 per cent of (sub)haloes that have the largest $V_{\rm scat}$ (\refeq{eq:Vpeak_scat}) in the UNIT simulation would not host ELGs. Although QSOs are the brightest object at $z>1$, their best-fitting $V_{\rm ceil}$ is inconsistent with zero, suggesting that not all haloes above a certain mass will be hosting a QSO. Their absence in the centre of massive (sub)haloes is consistent with the depletion of cold gas in this hot and dense environment there, leading to the quenching of ELG star formation and QSO black hole accretion. This agrees with the scenarios found in SAM studies and observations \citep[e.g.,][]{QSOincomplete2017,griffin2019,violeta_number_density}.

$v_{\rm smear}$ quantifies the effect that the redshift uncertainty has on the clustering. It can also be measured statistically and independently by the redshift difference $\Delta v$ of repeat observations (see Section~\ref{subsec:repeats}). The $\Delta v$ histogram of DESI tracers follows, in general, a Lorentzian profile with width $w_{\Delta v}$, instead of a single Gaussian profile as was found for BOSS/eBOSS galaxies. Thus, we have developed SHAM algorithms with a truncated Lorentzian profile for modelling the redshift uncertainty, i.e., $v_{\rm smear,L}$. The $\Delta v$ of LRG sub-samples in different redshift bins can be fitted well by a Lorentzian or a Gaussian profile and thus, we also develop a Gaussian $v_{\rm smear,G}$ for LRGs. The Lorentzian $v_{\rm smear, L}$ of LRGs is only consistent with $w_{\Delta v}$ at $0.4<z<1.1$. The clustering of LRG sub-samples in redshift bins actually suggests a preference for a Gaussian profile for the redshift uncertainty. Nevertheless, truncated Lorentzian and Gaussian functions provide the same SHAM clustering and consistent best-fitting $\sigma$ and $V_{\rm ceil}$. For QSOs, $v_{\rm smear,L}$ monotonically increases and deviates from $w_{\Delta v}$ at $z>1.5$. This is because the C~\textsc{iv} line used to determine QSO redshifts is affected by the velocity shifts of spectral lines that can vary between objects. Although the repeat observation cannot capture this feature,  its effect on the clustering will be modelled by SHAM $v_{\rm smear,L}$. This is consistent with the eBOSS QSO analysis \citep{Zarrouk2018}.

The satellite fraction $f_{\rm sat}$ of LRG and QSO samples is fixed to the number of subhaloes from the UNIT simulation included for the SHAM, given $\sigma$ and $V_{\rm ceil}$. Their $f_{\rm sat}$ decreases with redshift, following the evolution of the subhalo fraction in the simulations. For ELGs we use the $f_{\rm sat}$-SHAM, setting $f_{\rm sat}$ as a free parameter. The best-fitting $f_{\rm sat}$ for DESI ELGs is around 4 per cent. This low value is consistent with previous studies for strong \oii{} emitters~\citep{violeta_number_density,Gao2022}, but it is lower than the estimations for the total ELG samples \citep{favole_clustering_2016,eboss_elg_fsat,eboss_elg_sham}.

We provide the halo occupation distribution (HOD) measured from our best-fitting SHAM for LRGs, ELGs and QSOs from the One Percent Survey. The HOD of SHAM central LRGs reaches its peak at $\langle N\rangle=0.75\pm0.07$ and is consistent with an incomplete LRG pattern as found in \citep{Yuan2023}. The HOD for central SHAM ELGs is consistent with a star-forming HOD profile peaking at $\langle N_{\rm cen}\rangle=0.06\pm0.03$ and $M_{\rm vir}=10^{11.7}\Msun{}$, but we cannot exclude a Gaussian shape~\citep[e.g][]{eboss_elg_hod}. The HOD for SHAM central QSO also decreases after $M_{\rm vir}=10^{12.4}\Msun{}$ with $\langle N\rangle=0.016\pm0.001$. The HOD for all types of SHAM satellite galaxies is composed of two exponential functions with different slopes. The slope $\alpha$ in the massive halo end is $\alpha_{\rm LRG}={0.70}^{+0.01}_{-0.01}$, $\alpha_{\rm ELG}={0.76}^{+0.06}_{-0.07}$, and $\alpha_{\rm QSO}={0.73}^{+0.07}_{-0.07}$. They are smaller than the measurements from \absum HOD tests for LRGs but consistent with those from ELGs and QSOs \citep{Yuan2023,Rocher2023}. We shall point out that the decreasing halo occupation of centrals with respect to the halo mass for LRGs, ELGs and QSOs is a result of the simple $V_{\rm scat}$ truncation, i.e., the implementation of $V_{\rm ceil}$. Galaxy clustering produced by this profile is consistent with that from HOD models with other profiles. So we need more data with higher accuracy and physical models in SHAM/HOD to give a better description of this halo occupation incompleteness on the massive end. The cross-validation of the halo occupation number among hydrodynamical simulations, SAM, forward modelling, SHAM and HOD is planned for future work. This is because we have yet a consistent clustering measurement with the observation for all methods and the series of mock galaxies generated by those methods on the same simulation. 

We measure a mean parent halo mass of $\langle M_{\rm vir}\rangle=10^{13.16\pm0.01}\Msun{}$ for LRGs, $10^{11.90\pm0.06}\Msun{}$ for ELGs and $10^{12.66\pm0.45}\Msun{}$ for QSOs. For sub-samples at redshift bins, we obtain $\langle M_{\rm vir}\rangle$ that decreases with redshift for LRGs and QSOs, but not for ELGs. Meanwhile, the linear bias for each tracer increases with the redshift. Those results are consistent with the HOD measurement using the same tracers from the One Percent Survey in general. 

We also provide the SHAM-calibrated probability distribution of $V_{\rm peak}$ for LRGs, ELGs and QSOs at $0.8<z<1.1$. LRGs and QSOs are populated in (sub)haloes with a similar range of $V_{\rm peak}$, $\langle V_{\rm peak, LRG}\rangle=457\pm6\kms$ and $\langle V_{\rm peak, QSO}\rangle=346\pm8\kms$. The value for ELG (sub)haloes is smaller, $\langle V_{\rm peak, ELG}\rangle=159\pm4\kms$. This result will be useful for future multi-tracer studies.

SHAM algorithms that include the redshift uncertainty, massive-(sub)halo incompleteness and an adjustable satellite fraction work well in the single-tracer case, which can provide galaxy mocks for cosmological tests \citep[e.g.,][]{Su2023}. We plan to enhance this study in the future by implementing a multi-tracer SHAM method based on what we have learned from the present study.

\section*{Data Availability}
All the figures and the best-fitting mock galaxies of SHAM are available in Zenodo: \url{https://doi.org/10.5281/zenodo.7889632}.


\section*{Acknowledgements}
JY, CZ and JPK acknowledge support from the SNF 200020\_175751 and 200020\_207379 ``Cosmology with 3D Maps of the Universe" research grant. VGP is supported by the Atracci\'{o}n de Talento Contract no. 2019-T1/TIC-12702 granted by the Comunidad de Madrid in Spain and by the Ministerio de Ciencia e Innovaci\'{o}n (MICINN) under research grant PID2021-122603NB-C21. We would like to thank Risa Wechsler, Christophe~Y\`eche, Zheng Zheng, Charling Tao, Philip Mansfield, Hong Guo, Jeffrey Newman, Cheng Li, Haowen Zhang, Xiangyu Jin, Xi Kang and Svyatoslav Trusov for their helpful discussions. We also thank John Helly for providing us with the Millennium simulation.

This material is based upon work supported by the U.S. Department of Energy (DOE), Office of Science, Office of High-Energy Physics, under Contract No. DE–AC02–05CH11231, and by the National Energy Research Scientific Computing Center, a DOE Office of Science User Facility under the same contract. Additional support for DESI was provided by the U.S. National Science Foundation (NSF), Division of Astronomical Sciences under Contract No. AST-0950945 to the NSF’s National Optical-Infrared Astronomy Research Laboratory; the Science and Technology Facilities Council of the United Kingdom; the Gordon and Betty Moore Foundation; the Heising-Simons Foundation; the French Alternative Energies and Atomic Energy Commission (CEA); the National Council of Science and Technology of Mexico (CONACYT); the Ministry of Science and Innovation of Spain (MICINN), and by the DESI Member Institutions: \url{https://www.desi.lbl.gov/collaborating-institutions}. Any opinions, findings, and conclusions or recommendations expressed in this material are those of the author(s) and do not necessarily reflect the views of the U. S. National Science Foundation, the U. S. Department of Energy, or any of the listed funding agencies.

The authors are honoured to be permitted to conduct scientific research on Iolkam Du’ag (Kitt Peak), a mountain with particular significance to the Tohono O’odham Nation.


\bibliographystyle{mnras}
\bibliography{reference} 



\appendix
\section{4-parameter SHAM}
\begin{figure*}
    \includegraphics[width=\linewidth]{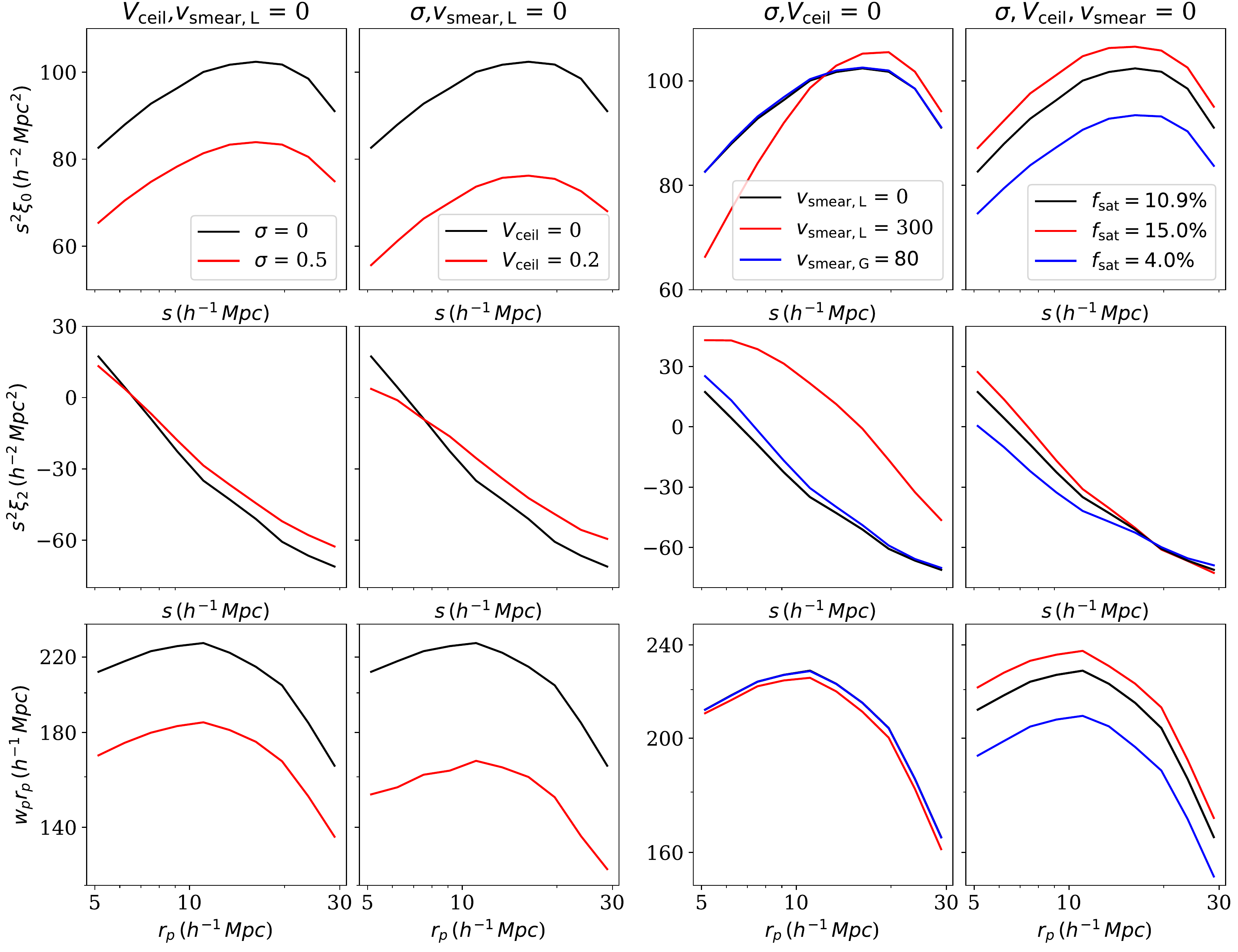}
    \caption{The impact of $\sigma$ (first column), $V_{\rm ceil}$ (second column), $v_{\rm smear}$ (third column), $f_{\rm sat}$ (fourth column) on the 2PCF monopole (first row), quadrupole (second row) and the projected 2PCF (third row). The standard sample is $\sigma,V_{\rm ceil},v_{\rm smear}=0$ with $f_{\rm sat}=10.9$ per cent in black lines. The first three columns show the 2PCF difference between the standard sample and samples with $\sigma=0.5$, $V_{\rm ceil}=0.2$ per cent or $v_{\rm smear,L}=300\kms$ (truncated at $2000\kms$) respectively while fixing the other two parameters, in red lines. Results of $v_{\rm smear}$-SHAM with a Gaussian profile $v_{\rm smear,G}=80\kms$ are also presented in the third column in blue lines. In the last column, we compare the 2PCF for SHAM galaxies with $f_{\rm sat}=15$ per cent (red lines) and $f_{\rm sat}=4$ per cent (blue lines) with that of the standard sample in $f_{\rm sat}$-SHAM with $\sigma,V_{\rm ceil},v_{\rm smear}=0$.
    }
    \label{fig:param impact}
\end{figure*}

\begin{figure}
    \includegraphics[width=\linewidth,scale=0.7]{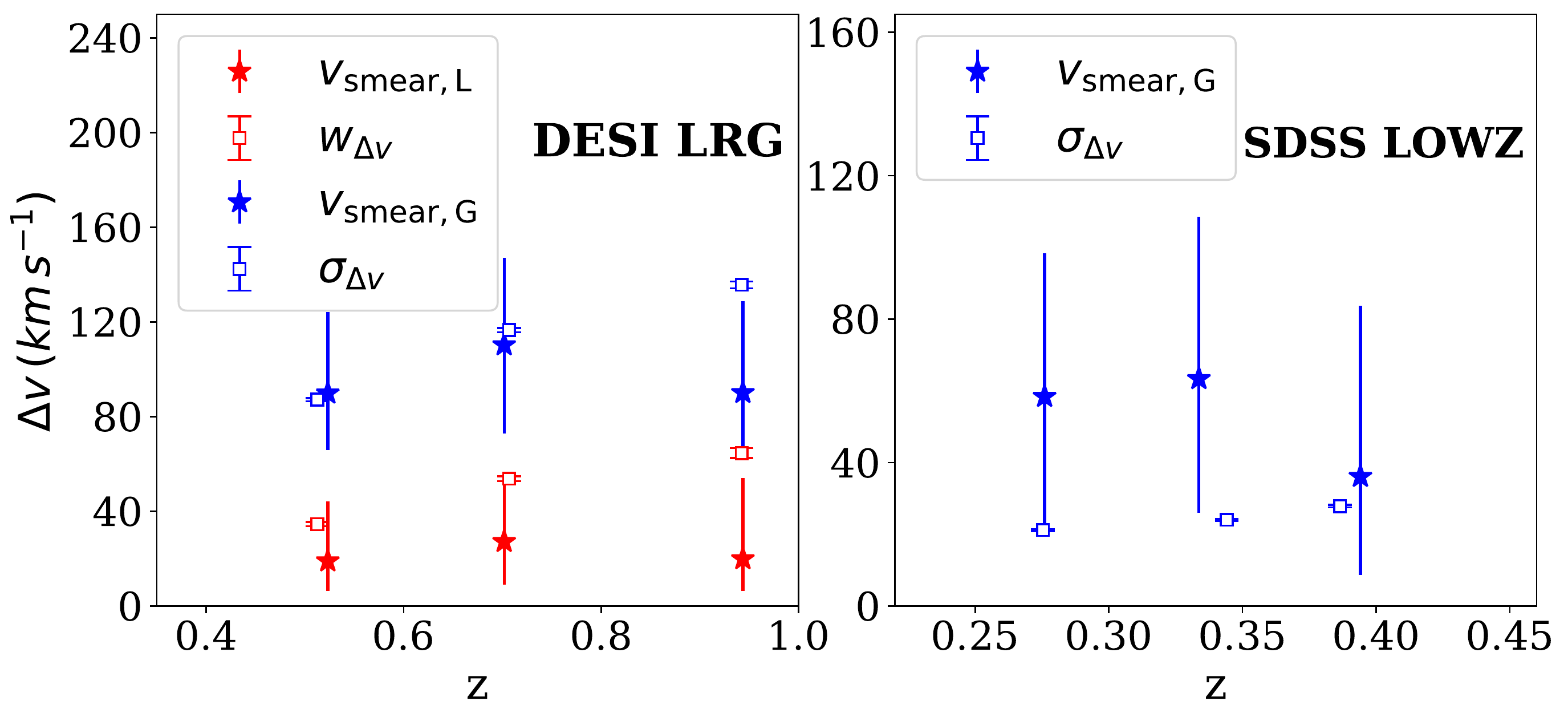}
    \caption{The redshift uncertainty measured from repeat observation (filled stars with error bars) and from SHAM (empty squares with error bars) for 4-parameter SHAM for LRGs (left panel) and SDSS-III LOWZ (right panel). Results for Lorentzian profiles are presented in red and those of Gaussian profiles are in blue. Note that the Gaussian measurements for DESI LRGs are shifted by $40\kms$ upwards. 
    }
    \label{fig:4param deltav}
\end{figure}
\begin{table*}
{
\begin{tabular}[c]{|c|c|c|c|c|c|c|c|c|c|}
{redshift}    & {$z_{\rm eff}$} & {$z_{\rm UNIT}$} &{ $V_{\rm eff}$}& {10$^4n_{\rm eff}$} & {$\sigma$} &  {$v_{\rm smear,G}$} & {$V_{\rm ceil}$} &{$f_{\rm sat}$}&{$\chi^2$/dof}\\
{range}    & {} & {} &{ $(h^{-3}\,{\rm Gpc}^{3})$}& {$(h^3\,{\rm Mpc}^{-3})$} & {}  & {$(\kms)$} &{$(\%)$}&{$(\%)$}& {}\\
\hline
\hline
{$0.2<z<0.43$}    & {0.3441}& {0.3337}  &{0.62}    & {2.95}&{$0.19^{+ 0.06}_{-0.06}$}&{$63^{+45}_{-37}$}&{$0.0045^{+0.0048}_{-0.0030}$} &{$13.55^{+2.07}_{-3.13}$}& {51/37}  \\
\hline
\hline
{$0.2< z< 0.33$} & {{0.2754}} & {{0.2760}} & {{0.29}}& {{3.37}} &{$0.15^{+ 0.09}_{-0.10}$}&{$58^{+40}_{-38}$}&{$0.0077^{+0.0060}_{-0.0048}$} &{$15.87^{+1.81}_{-3.11}$}& {32/37}  \\
\hline
{$0.33<z<0.43$} & {0.3865} & {0.3941} &{0.33} & {2.58}&{$0.27^{+ 0.05}_{-0.07}$}&{$36^{+48}_{-28}$}&{$0.0030^{+0.0038}_{-0.0021}$} &{$13.48^{+1.13}_{-2.66}$}& {50/37}  \\
\hline
\end{tabular}}
\caption{The same as Table~\ref{tab:3param result} but for 4-parameter SHAM with $\{\sigma,V_{\rm ceil}, v_{\rm smear},f_{\rm sat}\}$ applied on BOSS LOWZ clustering at $0.2<z<0.43$.}
\label{tab:SDSS tests}
\end{table*}

\label{appendix:LOWZ SHAM}
SHAM with 4 parameters $\{\sigma,V_{\rm ceil},v_{\rm smear},f_{\rm sat} \}$ is the inclusive version of $v_{\rm smear}$-SHAM and $f_{\rm sat}$-SHAM. 
\reffig{fig:param impact} provides the impact of those 4 parameters on the 2PCF monopole, quadrupole and projected 2PCF of UNIT-SHAM galaxies in the fitting range 5--30$\hmpc$. The `standard' clustering is obtained from $\sigma,V_{\rm ceil},v_{\rm smear}=0$ using $v_{\rm smear}$-SHAM. So its $f_{\rm sat}$ is derived from its SHAM catalogue. When we have a larger $\sigma$ as shown in the first column of \reffig{fig:param impact}, $\xi_0$ and $w_p$ decrease systematically and $\xi_2$ rotates counter-clock-wise with respect to a point at $s\sim 6\hmpc$. A larger $V_{\rm ceil}$ leads to a similar effect as shown in the second column, with the rotating point moving to $s\sim 7\hmpc$ in $\xi_2$. 
This is because both $\sigma$ and $V_{\rm ceil}$ control the mass range of (sub)haloes that can host model galaxies given a fixed $N_{\rm gal}$.

$v_{\rm smear}$ and $f_{\rm sat}$ are another pair of parameters that can change the velocity distribution along the line of sight, thus degenerated with each other. As presented in the third column of \reffig{fig:param impact}, increasing $v_{\rm smear}$ causes a larger $\xi_2$ on 5--30$\hmpc$. A Gaussian $v_{\rm smear,G}$ does not have a huge impact on $\xi_0$ and $w_p$, while a Lorentzian profile $v_{\rm smear,L}$ with a truncation in $2000\kms$ does influence $\xi_0$. Note that the clustering effect of a Lorentzian profile can be similar to that of a Gaussian profile as shown in Section~\ref{subsec:vsmear and deltav}. The difference in the clustering effect here should be attributed to both the $v_{\rm smear}$ value and truncation value. Nonetheless, those two symmetric $v_{\rm smear}$ profiles can only lead to an increasing $\xi_2$ at 5--30$\hmpc$ for SHAM galaxies compared to that of the `standard' sample. In contrast, we can decrease $\xi_2$ w.r.t. the `standard' one at similar scales by decreasing $f_{\rm sat}$. Note that those are the critical scales to reproduce the clustering of DESI ELG. Varying $f_{\rm sat}$ results in a systematical shift for $\xi_0$ and $w_p$ as well. 

We apply the 4-parameter SHAM to LRGs from the One Percent Survey for a consistency check with $v_{\rm smear}$ SHAM. We also perform a SHAM test for SDSS-III BOSS LOWZ samples, trying to resolve the overestimation of redshift uncertainty by $v_{\rm smear}$-SHAM found in \citetalias{mine}. In \reffig{fig:4param deltav}, we present the $v_{\rm smear}$ from the best-fitting 4-parameter SHAM together with $w_{\Delta v}$ and $\sigma_{\Delta v}$ from the repeat observations for DESI LRGs (left) and LOWZ LRG samples (right). Comparing with \reffig{fig:deltav evolution}, $v_{\rm smear}$ values systematically shift to smaller values. The satellite fraction of 4-parameter SHAM is consistent with that of $v_{\rm smear}$-SHAM, except for LRGs at $0.8<z<1.1$ with Gaussian profile, for which the $v_{\rm smear,G}$ result becomes inconsistent with $\sigma_{\Delta v}$ of repeat observations. As we have a reliable statistical measurement of LRG redshift uncertainty, it means that the satellite fraction estimation might be biased by the redshift uncertainty.

For the LOWZ 4-parameter SHAM study, the basic information of LOWZ observation, including UNIT simulations and the best-fitting SHAM results is presented in Table~\ref{tab:SDSS tests}. Our best-fitting $v_{\rm smear}$ are consistent with $\sigma_{\Delta v}$, resolving the discrepancy shown in Figure \textcolor{blue}{6} of \citetalias{mine}. The $v_{\rm smear}$--$f_{\rm sat}$ degeneracy enables $v_{\rm smear}$ to decrease to the observed uncertainty level by increasing its satellite fraction. Meanwhile, our best-fitting $f_{\rm sat}$ is consistent with the LOWZ HOD $f_{\rm sat}=12\pm2$ per cent \citep{2013LOWZsat}. 

\section{Posterior Contours}
\label{appendix:Posterior Contours}
\begin{figure}
    \centering
    \includegraphics[width=\linewidth]{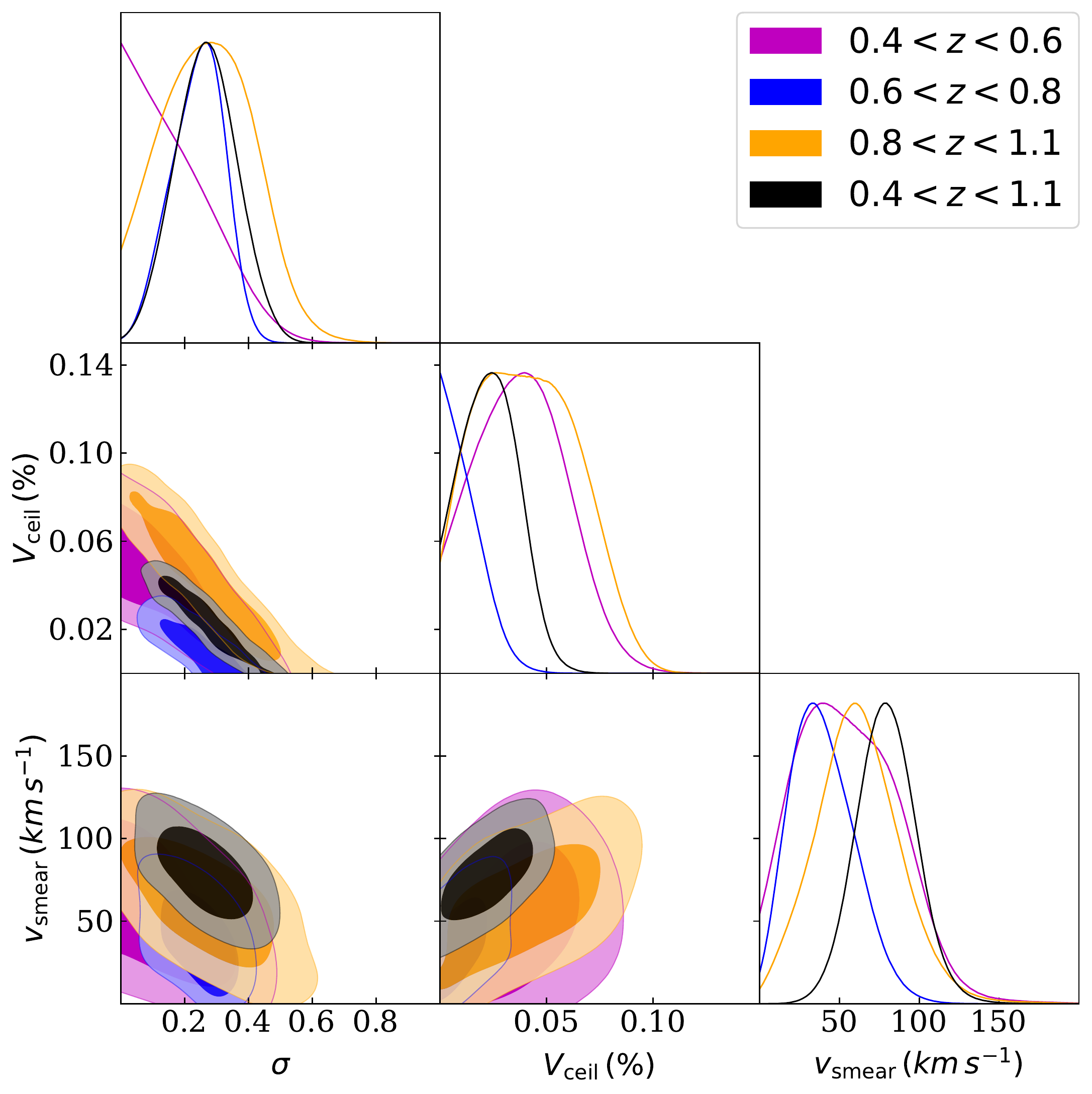}
    \caption{The posterior corner plot for LRGs at $0.4<z<0.6$ (magenta), $0.6<z<0.8$ (blue), $0.8<z<1.1$ (orange), and $0.4<z<1.1$ (grey). The parameters are $\sigma$, $V_{\rm ceil}$, $v_{\rm smear}$. }  
    \label{lrg posterior}
\end{figure}

\begin{figure}
    \centering
    \includegraphics[width=\linewidth]{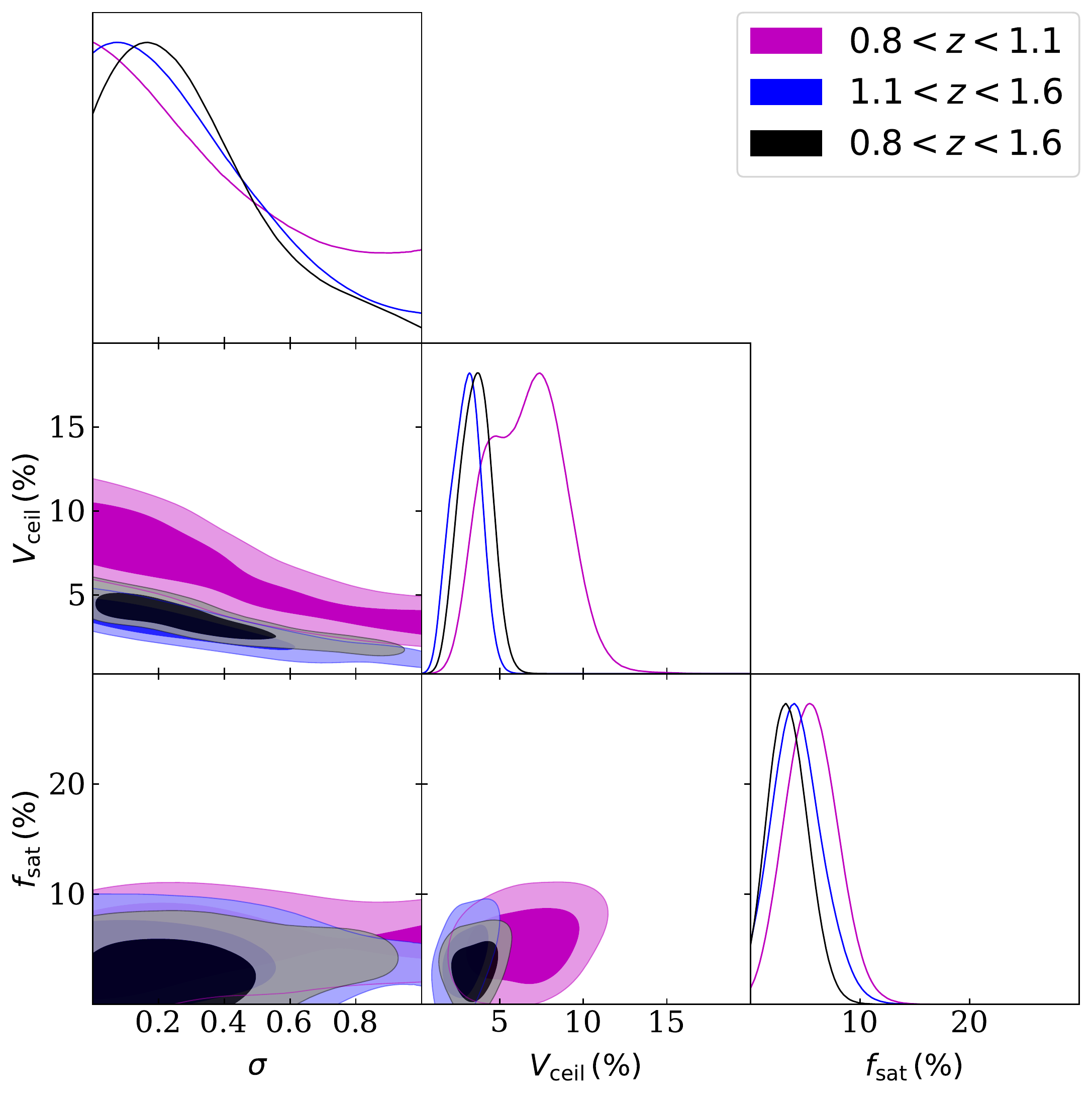}
    \caption{The posterior corner plot for ELGs at $0.8<z<1.1$ (magenta), $1.1<z<1.6$ (blue) and $0.8<z<1.6$ (grey). The parameters are  $\sigma$, $V_{\rm ceil}$, $f_{\rm sat}$. }  
    \label{elg posterior}
\end{figure}

\begin{figure}
    \centering
    \includegraphics[width=\linewidth]{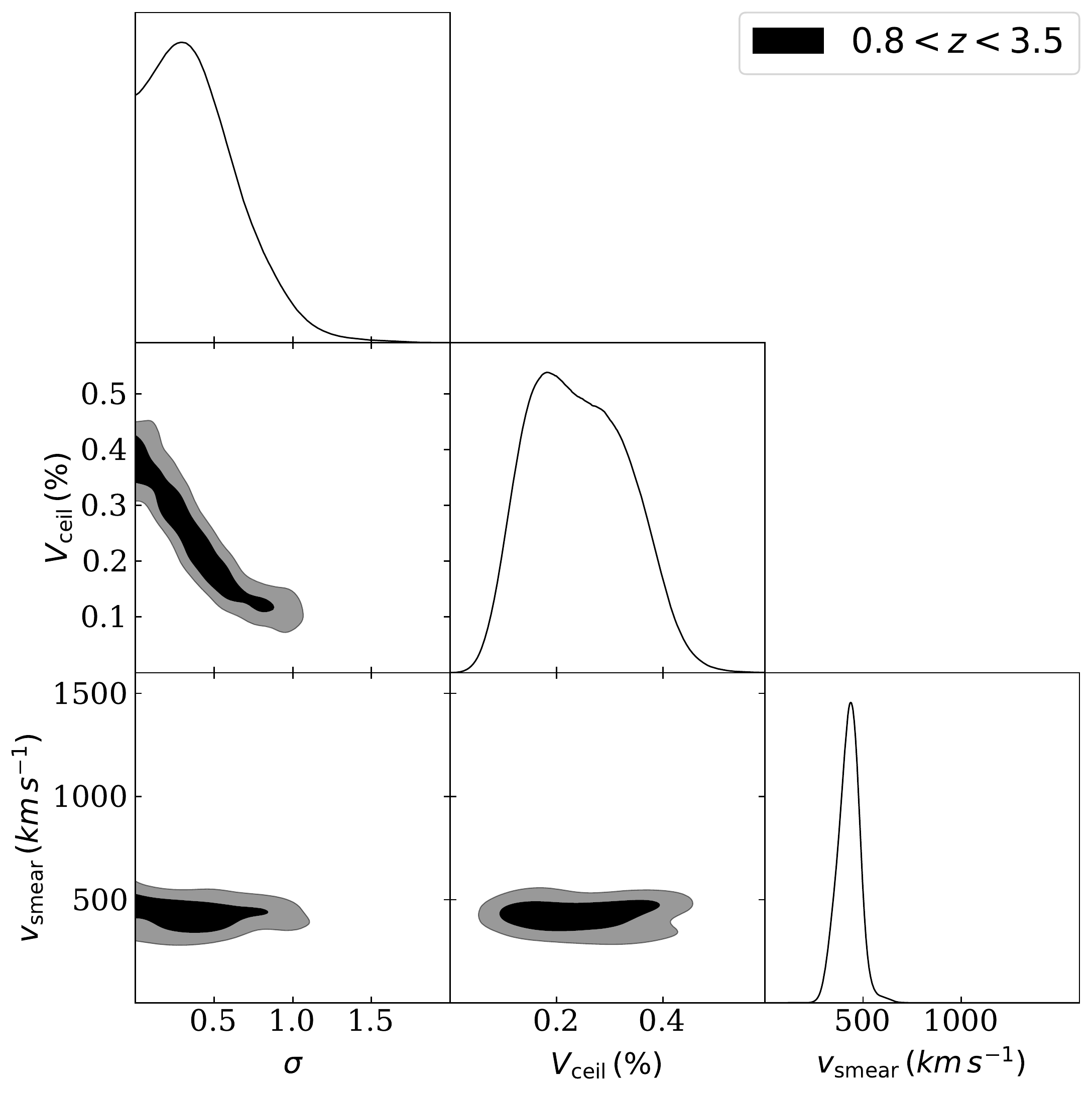}
    \caption{The same as \reffig{lrg posterior} but for QSOs at $0.8<z<3.5$.}
    \label{qso bulk posterior}
\end{figure}

\begin{figure}
    \centering
    \includegraphics[width=\linewidth]{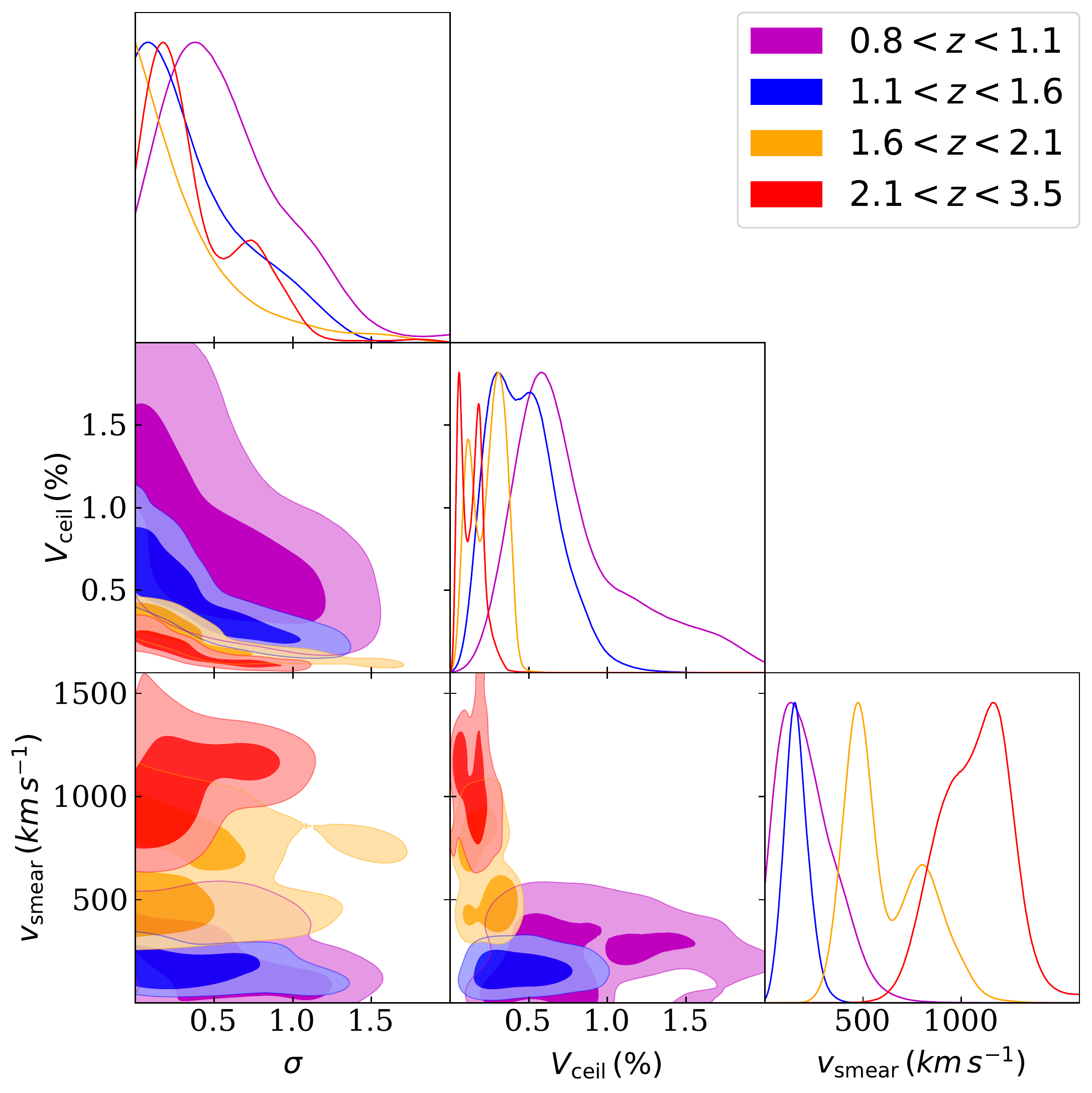}
    \caption{The same as \reffig{lrg posterior} but for QSOs at $0.8<z<1.1$ (magenta), $1.1<z<1.6$ (blue), $1.6<z<2.1$ (orange), and $2.1<z<3.5$ (red). }  
    \label{qso posterior}
\end{figure}

In Figures~\ref{lrg posterior}--\ref{qso posterior}, we present the posteriors of the fittings from Table~\ref{tab:3param result}, i.e., those of $v_{\rm smear}$-SHAM for LRG samples and QSOs, and $f_{\rm sat}$-SHAM for ELG samples. The $v_{\rm smear}$ profile here is Lorentzian. They are plotted using \textsc{getdist} \citep{2019arXiv191013970L}. All fittings have converged and are ended by the nested sampling automatically. 


\section{Reproduced \texorpdfstring{$w_p$}{wp}}
\label{appendix:reproduced wp}
\begin{figure}
	\includegraphics[width=\columnwidth]{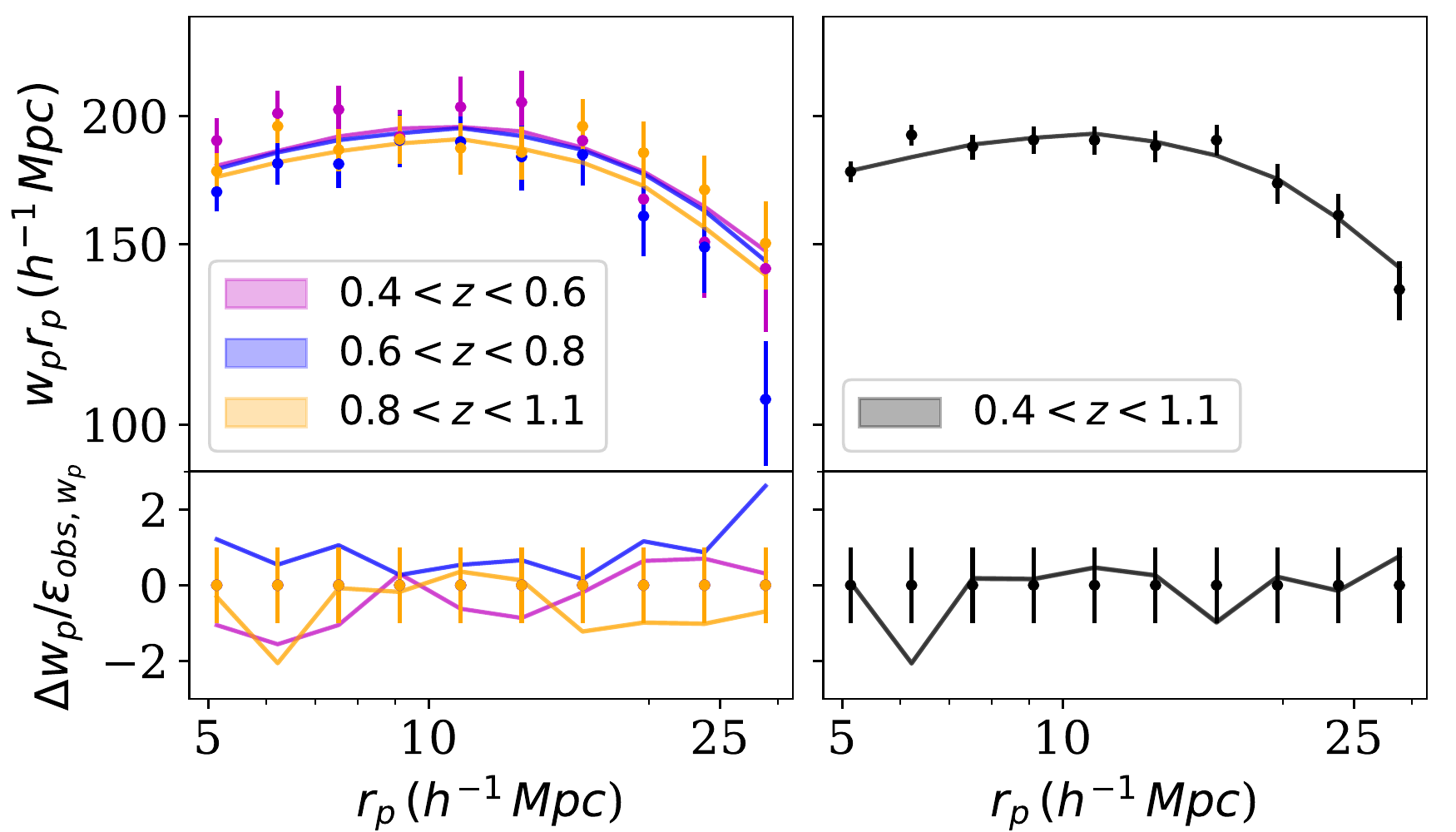}
    \caption{The projected 2PCFs $w_pr_p$ of observed LRGs (filled circles with  error bars) and those of the best-fitting SHAM galaxies (solid lines with shades) with $\pi_{\rm max}=30\hmpc$. The direct comparison and residuals rescaled by the observed error bars are presented in the first and the second row respectively. The samples on the left are LRGs at different redshift bins, while the sample on the right is the total sample. }
    \label{fig:lrg wp}
\end{figure}

\begin{figure}
	\includegraphics[width=\columnwidth]{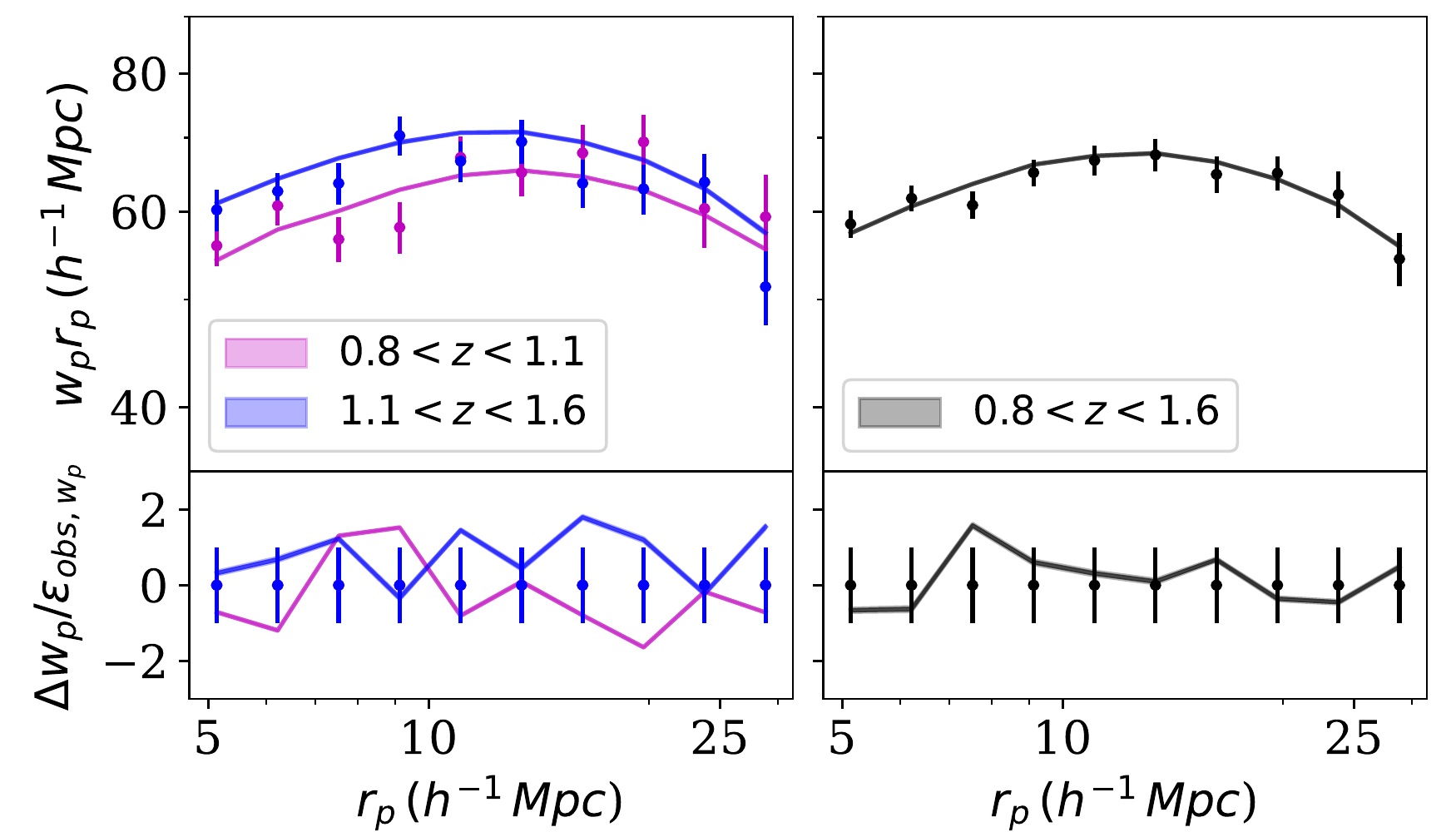}
    \caption{Same as \reffig{fig:lrg wp} but for ELGs.}
    \label{fig:elg wp}
\end{figure}

\begin{figure}
	\includegraphics[width=\columnwidth]{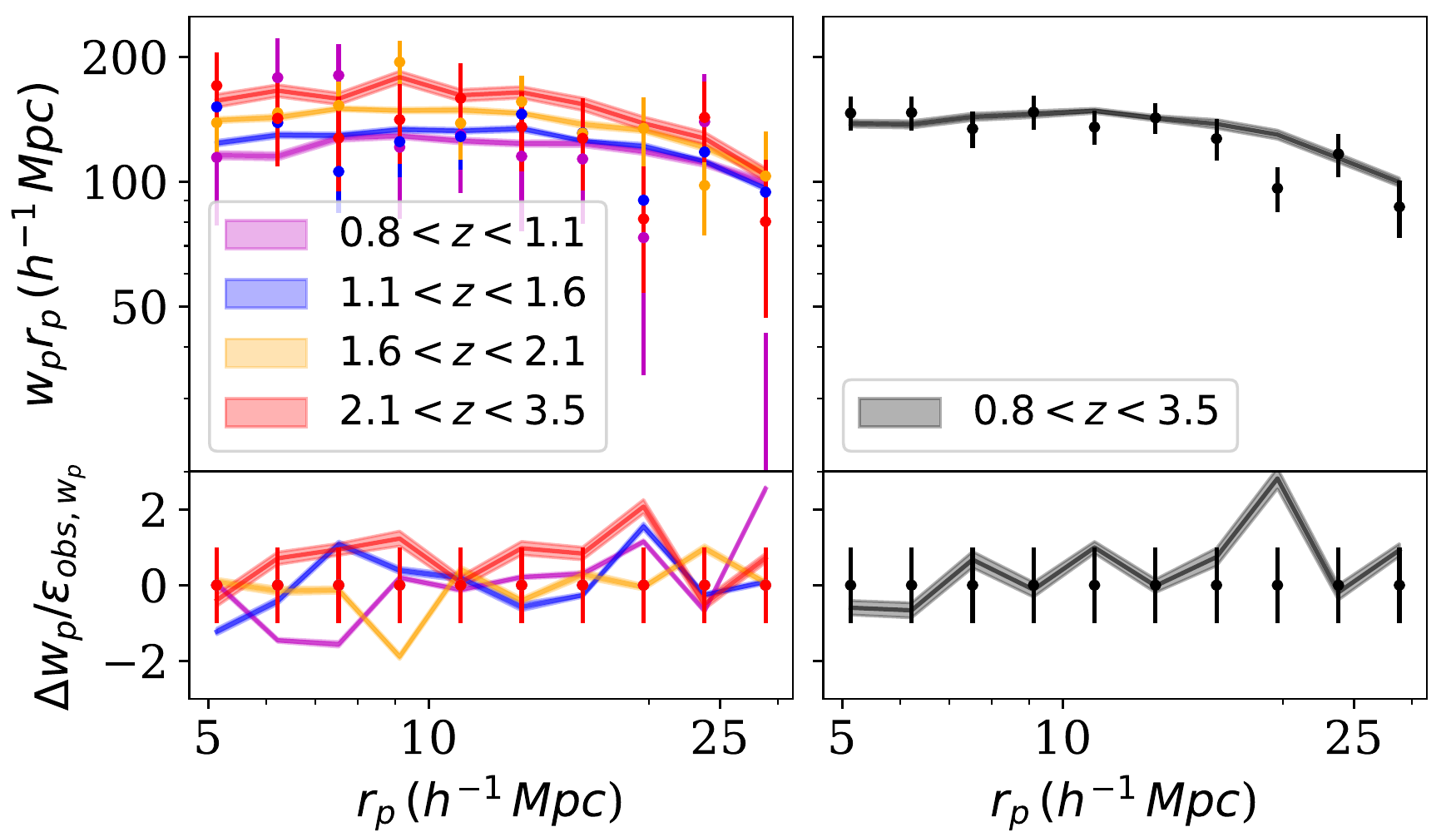}
    \caption{Same as \reffig{fig:lrg wp} but for QSOs.}
    \label{fig:qso wp}
\end{figure}

We provide the comparison between mock galaxies of SHAM and observations for the projected 2PCF $w_pr_p$ at 5--30$\hmpc$ in Figures~\ref{fig:lrg wp}--\ref{fig:qso wp}. The mock galaxies are constructed using the parameter set of SHAM that corresponds to the minimum $\chi^2$. The $\pi_{\rm max}$ of their $w_p$ are $30\hmpc$ (Section~\ref{subsec:2pcf}). The best-fitting reduced $\chi^2$ of LRG at $0.8<z<1.1$ is larger than 1.5, but the reproduced $w_p$ agrees with the observation. So mock galaxies of SHAM for this sample are still a good description of the observed clustering. 

\begin{figure}
    \includegraphics[width=\columnwidth]{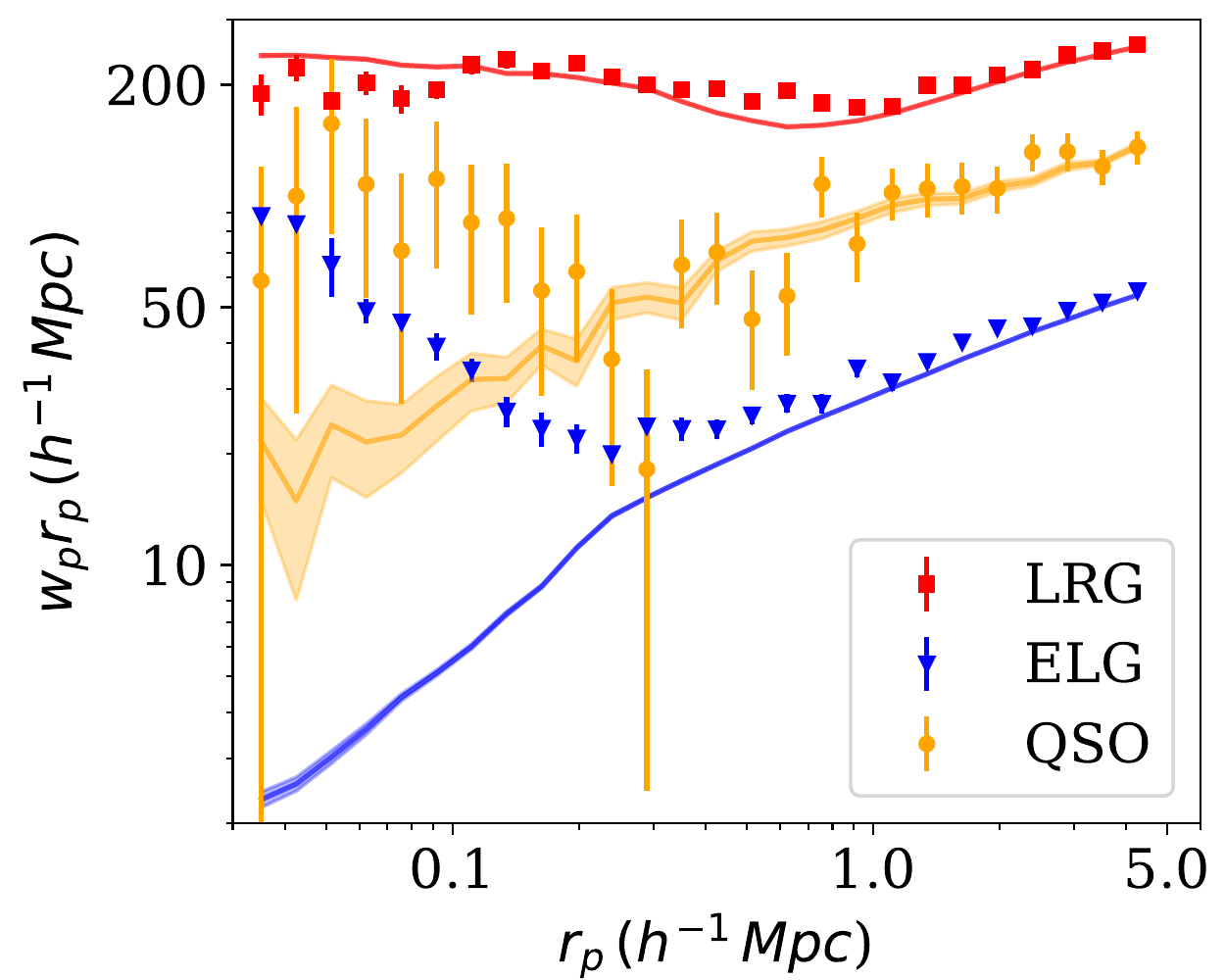}
    \caption{The projected 2PCF $w_pr_p$ at 0.01--5$\hmpc$ of observations (filled error bars) and the predictions from the best-fitting SHAM (solid lines with shades). Among the filled error bars, red squares are for LRGs, blue triangles are for ELGs and yellow circles are for QSOs. The colour of the solid lines with shades is the same as the colour of the corresponding error bars. The $w_pr_p$ of LRGs is vertically shifted.}
    \label{fig:wp smallscale}
\end{figure}

We note the disagreement on scales smaller than 1$\hmpc$ between the observation and the prediction of the best-fitting SHAM galaxy mocks in general as illustrated in \reffig{fig:wp smallscale}. This is probably due to the over-disruption or over-merging of subhaloes in $N$-body simulations \citep[e.g.,][]{subhalo_disrupt2018,subhaloe_disrupt_2_2018,Behroozi2019}. In addition to that, the uprising small-scale $w_pr_p$ of DESI ELGs may have physical explanations as shown in \citet{Rocher2023}.
\begin{figure*} 
    \centering
	\includegraphics[width=\linewidth]{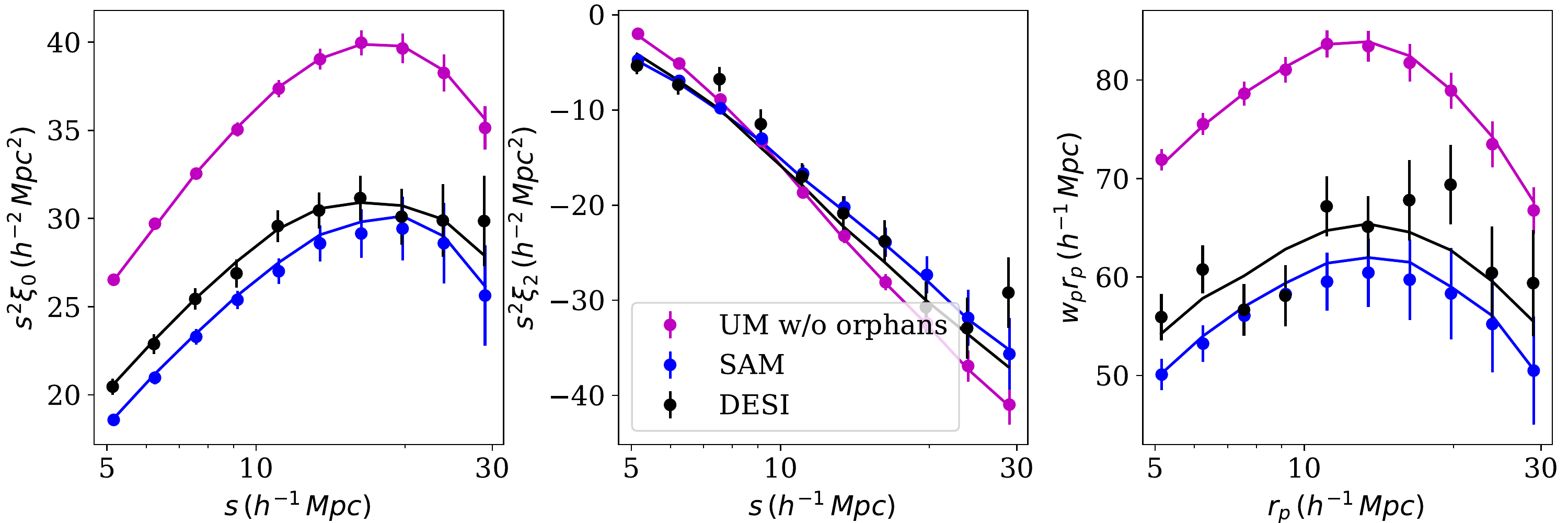}
    \caption{The 2PCF of UniverseMachine ELGs (magenta circles with error bars) at $z=0.9436$, SAM ELGs (blue circles with error bars) at $z=0.99$, and DESI ELGs (black circles with error bars) with $z_{\rm eff}=0.9565$, and their corresponding best-fitting SHAM galaxies. The 2PCF monopoles, quadrupole and projected 2PCF are presented in the left, middle and right panels respectively.}
    \label{fig:elg comparison}
\end{figure*}



\section{Is Satellite Fraction Biased?}
\label{appendix:fsat bias}
To validate our satellite fraction in $f_{\rm sat}$-SHAM measurement, we use other galaxy mocks as observations. They are constructed in various galaxy--halo models and have the same definition of satellites as our SHAM, i.e., galaxies residing in subhaloes. By implementing $f_{\rm sat}$-SHAM on the same $N$-body simulations as those model galaxies and fitting the 2PCF monopole and quadrupole of those modelled galaxies on 5--30$\hmpc$, $f_{\rm sat}$ of the best-fitting SHAM is expected to be consistent with the true value of those galaxy mocks. The covariance matrices we used here are calculated with jackknife subsamples of other mock galaxies produced by \textsc{pycorr}. 

The first set of galaxies is from a DESI-like ELG catalogue established with SAM \citep[][SAM ELG hereafter]{Millennium_WMAP,violeta_number_density} at redshift $z=0.99$ with $f_{\rm sat}=4.14$ per cent. The corresponding $N$-body simulation is \textsc{Millennium}\footnote{\url{https://virgodb.dur.ac.uk:8443/Millennium/Help?page=databases/gonzalez2014a/mr7}} with the WMAP7 cosmology \citep{Millennium2005}. As there is no $V_{\rm peak}$ in this simulation, we use $V_{\rm max}$ for $f_{\rm sat}$-SHAM. Another sample of star-forming galaxies from UniverseMachine \citep[][UniverseMachine ELG hereafter]{Behroozi2019} at $z=0.9436$ with star-formation rate larger than $10^{1.1}\,\rm \textup{M}_\odot yr^{-1}$ and $10^{9.6}<M_*<10^{11}\,\textup{M}_\odot$. This galaxy catalogue is based on the snapshot of \textsc{MultiDark} MDPL2 simulation \citep{MDPL2_2012} at the same redshift\footnote{\url{https://www.cosmosim.org/metadata/mdpl2/}}. Orphan galaxies \citep{Campbell2018, Behroozi2019} are removed from this sample to ensure a fair comparison to our SHAM-reproduced results as \textsc{MultiDark} simulations do not include this reconstruction on subhaloes by default. Finally, the star-forming galaxies from UniverseMachine are downsampled to have $n_{\rm gal}=10.28\times10^{-4}\,\rm Mpc^{-3}$$h^3$, 1.8 per cent smaller than that of DESI ELGs at $0.8<z<1.1$. 

\reffig{fig:elg comparison} shows the clustering of SAM ELGs (blue dots with error bars), UniverseMachine ELGs (magenta dots with error bars) and DESI ELGs (black dots with error bars), and the clustering of the best-fitting SHAM galaxies (lines with corresponding colours). Our $f_{\rm sat}$-SHAM can describe the 2PCF multipoles for SAM ELGs and UniverseMachine ELGs and reproduce their projected 2PCF on 5--30$\hmpc$. The best-fitting $f_{\rm sat}$ for SAM ELGs is $11.5^{+1.46}_{-1.19}$ per cent, which is a 6-$\sigma$ overstimation. Meanwhile, that for UniverseMachine ELGs is $12.60^{+0.76}_{-0.82}$ per cent, consistent with the true value. The inconsistency in the estimation of the satellite fraction among different galaxy--halo models needs further discussion in future studies.

\bsp	
\label{lastpage}
\end{document}